\documentclass[preprint,11pt,authoryear]{elsarticle}

\usepackage{amsmath}
\usepackage{amssymb}
\usepackage{soul}
\usepackage[T1]{fontenc}
\usepackage[utf8]{inputenc}
\usepackage{hyperref}

\usepackage[margin=2.5cm,footskip=2cm]{geometry}
\usepackage{booktabs}
\usepackage{longtable}
\usepackage{algorithm}
\usepackage[noend]{algpseudocode}
\newcommand\numberthis{\addtocounter{equation}{1}\tag{\theequation}}
\journal{a journal}

\begin{document}

\begin{frontmatter}
\title{Guiding large-scale management of invasive species using network metrics}

	\author[a]{Jaime Ashander\corref{1}\fnref{2}}
	\cortext[1]{To whom correspondence should be addressed.}
	\fntext[2]{Current address: U.S. Geological Survey, Eastern Ecological Science Center.}
	\ead{jashander$@$usgs.gov}
\author[c,a]{Kailin Kroetz}
\author[a]{Rebecca Epanchin-Niell}
\author[d]{Nicholas B. D. Phelps}
\author[e]{Robert G. Haight}
\author[b]{Laura E. Dee}

  \affiliation[a]{Resources for the Future, 1616 P Street NW, Suite 600, Washington, DC 20036}
  \affiliation[b]{University of Colorado, Boulder, Department of Ecology and Evolutionary Biology, Boulder, CO USA 80309}
  \affiliation[c]{Arizona State University, School of Sustainability, PO Box 87550, Tempe, AZ USA 85287-5502}
  \affiliation[d]{University of Minnesota, College of Food, Agricultural, and Natural Resource Sciences, Department of Fisheries, Wildlife, and Conservation Biology, 2003 Upper Buford Cir., St. Paul, MN 55108}
  \affiliation[e]{USDA Forest Service, Northern Research Station, St. Paul, MN 55108, USA}


%%%%%%%%%%%%%%
%% ABSTRACT %%
%%%%%%%%%%%%%%

\begin{abstract}
Complex socio-environmental interdependencies drive biological invasions, causing damages across large spatial scales. For widespread invasions, targeting of management activities based on optimization approaches may fail due to computational or data constraints. Here we evaluate an alternative approach that embraces complexity by representing the invasion as a network and using network structure to inform management locations. We compare optimal versus network-guided invasive species management at a landscape-scale, considering siting of boat decontamination stations targeting 1.6 million boater movements among 9,182 lakes in Minnesota, USA. Studying performance for 58 counties, we find that when full information is known on invasion status and boater movements, the best-performing network-guided metric achieves a median and lower quartile performance of 100\% of optimal. We also find that performance remains relatively high using different network metrics or with less information (median above 80\% and lower quartile above 60\% of optimal for most metrics), but is more variable, particularly at the lower quartile. Additionally, performance is generally stable across counties with varying lake counts, suggesting viability for large-scale invasion management.
\end{abstract}

%\dates{This manuscript was compiled on \today}

\end{frontmatter}

Complex socio-environmental interdependencies drive biological invasions at both regional and global scales \cite{banks_role_2015,epanchin-niell_complex_landscapes_2009}.
These invasions damage ecosystems worldwide, threatening biodiversity and ecosystem services \cite{charles_impacts_2007,gallardo_global_2016} and causing
annual losses of over \$46B USD and rising \cite{diagne_high_2021}. 
Over the coming decades, increases in global interconnections and associated trade
are expected to drive many more invasions \cite{sardain2019global}. 

Widespread biological invasions, which cover large spatial scales, are a major environmental crisis whose 
long-term solution will require advancing beyond existing tools for managing established invasions
\cite{epanchin-niell_controlling_2010}. 
Novel, approximate methods for controlling spread at large scales 
are needed because existing accepted methods for finding exact optima for these spatial resource allocation problems 
(e.g., using optimal control, mixed integer linear and nonlinear programming \cite[e.g.,][]{chades_general_2011,epanchin-niell_optimal_2012,epanchin-niell_individual_2015,aadland2015spatial,baker2017target,bushaj2021optimizing,fischer2021managing}) cannot scale efficiently to large systems \cite{buyuktahtakin_review_2018}. 
These current methods are also often hindered by the challenge of limited data.
Both the distribution of invasive species and the socio-ecological processes driving spread are at best partially-observed even in established invasions \cite{epanchin-niell_controlling_2010,epanchin-niell_economics_2017}.

Network approaches are a promising way to analyze complex socio-environmental systems \cite{bodin2019improving}. 
Over the past decade, network science has made advances in understanding how to optimally control spread in complex networks, where the ``network protection problem'' is fundamental with applications in information security, epidemiology, politics, and marketing \cite{nowzari_analysis_2016}.
Recognition that optimization approaches for network protection cannot scale to large network sizes due to problem complexity has led to intensive research effort focused on ``network-guided management": heuristic and approximate methods for solving this problem \cite{newman_spread_2002,kempe_maximizing_2003,pastor-satorras_immunization_2002, pastor-satorras_epidemic_2015} including using network metrics to prioritize management actions \cite{holme_attack_2002}.

Despite clear parallels between network protection and invasive species management, and network approaches becoming more commonly applied to invasive species or socio-environmental systems (SESs), network-guided management has not been widely adopted for managing biological invasions. Specifically, the parallel exists because an invasion can be represented as a network, with sites that an invasive species can occupy representing nodes of a network and edges representing pathways of spread (e.g., via human movement using gravity models \cite{muirhead_development_2005}, or via biological dispersal using habitat connectivity models \cite{de2018predicting}). 
As such, there is growing interest among researchers and practitioners in using network characteristics to understand invasions. Network characteristics affect spread both at landscape scales \cite{minor_graph-theory_2008} and continental or global scales, where invasions are most often mediated by human movement and trade \cite{banks_role_2015}. For example, consistent with network theory, experimental work reveals that spread is facilitated by network hubs, or central patches through which many dispersal pathways flow, and hindered by more clustered network structures \cite{moreljournel_its_2018}.  However, these insights are not widely used to inform management. One explanation for this is that rigorous comparisons between network-guided management and accepted methods to obtain optimal solutions have been done only for newly established invasions in small systems---a situation where management based on network structure performs poorly \cite{chades_general_2011}. Despite their poor performance for newly established invasions \cite{chades_general_2011}, network-guided management has been suggested for landscape-scale management of spreading species \cite{perry_using_2017}, including aquatic invasive species (AIS) that spread via human trade and movement \cite{kvistad2019network}.
To understand if network approaches provide value over accepted methods for optimal management of \textit{large} networks with established invasions, however, rigorous comparisons are needed but have not yet been done. 

Here we aim to bridge the gap between advances in network science on spread prevention in complex networks and real-world management of invasions in SESs.
Specifically, our approach connects recent inquiry into using network structure to guide management of invasive species with existing understanding of optimal management of species invasions.
We focus on a case of managing a large-scale invasion via allocation of inspection and decontamination stations (``inspection stations'' hereafter) across a network, with the objective of minimizing the number of uninspected but potentially infective connections, a common problem in invasive species management
\cite{haight2021paper,fischer2021managing}.
We evaluate performance of management based on network structure (i.e., centrality metrics, see ``Analytical framework''; \cite{holme_attack_2002}) relative to the optimal management from an integer linear programming solution \cite{haight2021paper,fischer2021managing}.
Since managers are often constrained by knowledge of infestation status and processes (e.g., current invasion distribution, dispersal patterns), we also examine how performance degrades with reduced information on invasion status or spread magnitude, relative to the optimal solution which requires full information.
Additionally, we characterize performance across a range of budgets.
Thus, our analysis advances knowledge about the outstanding question of whether network metrics can effectively guide management and whether some metrics could work better than others in more information rich versus information poor management settings.

%%%%%%%%%%%%%%%%%%%%%%%%%%
%% APPLYING AND TESTING %%
%%%%%%%%%%%%%%%%%%%%%%%%%%

\section*{Applying and testing network-guided management}
\subsection*{Study system and management objective}
We focus on the large-scale management problem of preventing zebra mussel ({\it Dreissena polymorpha}) spread in Minnesota, USA (MN).
Zebra mussels are one of the most costly invasive species in the US,  causing changes to ecosystem processes \cite{mceachran2019stable}, extirpating native mussel species \cite{karatayev2002impacts}, and damaging infrastructure \cite{prescott2013impact}.
In MN, zebra mussels are a prohibited invasive species, yet have spread to more than 270 lakes/rivers since their first report in 1989, with more than half of those infestations occurring since 2016 (see \cite{MNDNR2020} \& Supplemental Information (SI) Section S1 for more details).
This rapid spread has been largely human-mediated, facilitated by movement of boats and equipment \cite{kanankege2018probability, mallez2018dispersal}. 
The objective we examine, minimizing short-term spread via optimal location of inspection stations at lakes, is implicitly aimed at achieving the more fundamental goal of reducing long-term damages from invasion spread \cite{MNDNR2020}.
Given that significant state funding is delegated to the county level (\$10M annually for AIS prevention; \cite{MNDNR2020}),
decisions on where to locate most inspection stations are made by county managers \cite{MNDNR2020}. 
Indeed, two co-authors (Phelps and Haight) were approached by state and county managers to both characterize boat movements \cite{kao2021paper} and develop decision support tools for cost-effective county-level management of watercraft inspection programs \cite{haight2021paper}.
 Here, for each of the 58 counties in MN with more than 10 lakes, we examine the performance of network-guided approaches relative to an optimal solution obtained via integer linear programming (see Methods). Our set-up allows us to also examine how performance varies across counties that differ in system size (number of nodes).

\subsection*{Analytical framework}

\begin{figure}
\includegraphics{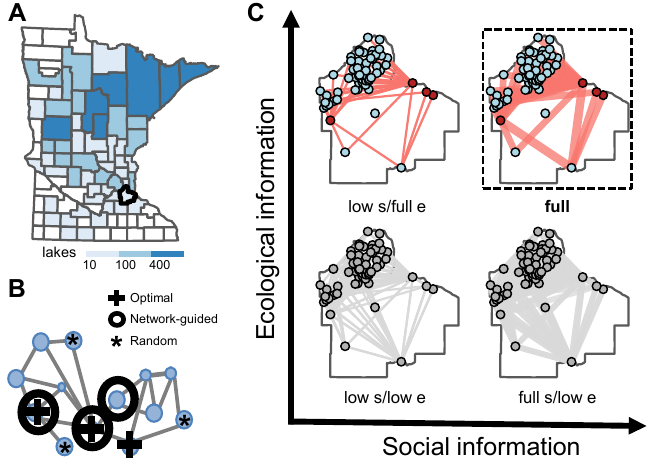}
    \caption{Our analytical approach:
    (A) 
    Networks of boat movement are constructed at the county level for Minnesota counties with 10 or more lakes ($n=58$; Table S1)  shaded here by the number of lakes in the county, which corresponds to network size. Dakota County, MN (bold outline) is used for illustration in part C of the figure.
(B) Network-guided management (placement of inspection stations) using heuristics based on centrality metrics (see Table~\ref{tab:metrics}) is compared both to optimal inspection (using integer-programming) and random management.
    (C) In addition to examining performance of network-guided management with full information (dashed outline), we evaluate performance in three cases with only partial information on either or both the boater movement network (``social information'') and the invasion status of each lake (``ecological information'').
See Table~\ref{tab:info} for more detailed definitions of full and low social and ecological information.
    }\label{fig:overview}
\end{figure}

Our analytical approach centers around the optimal placement of boat inspection stations 
on a network of infested and uninfested lakes connected by boater movements. 
We compare the optimal solution from integer linear programming to {\em network metric-guided} 
management with sites prioritized based on the structure of the boater movement network. Our general approach involves six key steps.

(1) \textit{Construct the network.}  Given that the primary mode of zebra mussel spread is through human movement (of boats)  \cite{kanankege2018probability}, we define a network with lakes that are infested or not as nodes, and the number of boats moving between lakes as weighted, directed edges that mediate spread of infestation (see Methods \& SI Section S2; \cite{kao2020}).
Within the overall network there is an ``infested subnetwork''---the network of potentially infective movements from infested lakes to uninfested lakes (Fig.~\ref{fig:overview}B).
We construct a network for each county with 10 or more lakes ($n=58$; Table S1). 

(2) \textit{Compute the optimal inspection strategy for each county.} The management objective is to maximize the inspections of potentially-infective boats, i.e., boats moving from infested to uninfested lakes. 
We focus on county-level management, which controls 90\% of the inspection stations in MN.
County managers decide where to locate inspection stations given a fixed budget, often with minimal coordination between counties or with the state.  Thus, we solve for each county's optimal inspection pattern, subject to a fixed budget, using integer linear programming (see Methods) where the objective is to maximize the number of potentially infective boats inspected (see SI Section S3).

(3) \textit{Compute network-guided inspection strategies for each county.}
We evaluate management strategies based on several centrality metrics that are measures of network structure (see Table~\ref{tab:metrics}). These include two strategies used in network protection contexts for the objective of reducing long-term spread: prioritizing based on highest degree or betweenness centrality \cite{holme_attack_2002}.
These metrics also have been identified as useful predictors of node importance for invasive species spread \cite{moreljournel_its_2018,banks_role_2015}.
We also propose and apply a novel metric that combines the Hub and Authority Scores of Kleinberg \cite{kleinberg1998authoritative}; our hub+authority score (H+A) favors nodes that are {\it either} sources of infestation or targets of infestation on a directed network.
In total, we examine five heuristic network-guided strategies (see SI Sections S4 \& S5), presenting results in the main text for three focal strategies that prioritize nodes with highest network centrality (see Table~\ref{tab:metrics} and Methods). 
 
To evaluate management strategies based on centrality metrics, we rank the lakes (nodes) in each county according to each centrality metric, and place interventions in order of priority based on these ranks until the budget runs out. We calculate the \textit{relative performance} of these strategies (Fig.~\ref{fig:overview}B), i.e., the proportion of potentially infective boats inspected compared to optimal inspection (See SI Section S7).

(4) \textit{Construct cases with reduced ecological and social information.}
The optimal solution provides an upper bound on the performance of network-guided management, but because data are a key constraint in managing large, complex socio-ecological systems, we seek insight into whether network-guided management can perform well when planners use network-guided approaches with less detailed social or ecological information.
We focus on four cases with varying levels of social and ecological information (Fig~\ref{fig:overview}C and Table~\ref{tab:info}): 
full social and full ecological information;
full social and low ecological;
low social and full ecological;
low social and low ecological. 
In both full and low social information cases, the directed topology of the network is known.
With full social information, the edge weights are known, and a weighted edge from one lake to another represents the number of boats moving from that lake to the other.
With low social information, the presence or absence of edges between two lakes are known. An edge from one lake to another represents at least one boat movement from that lake to the other, while no edge represents zero boat movements.
With full ecological information, lake infestation status is known.
Combined with the directed topology, this means the ``infested subnetwork''---the network consisting of only potentially infective movements between infested lakes and uninfested lakes--- is known with full ecological information.
With low ecological information, only the position of each lake in the network is known, not its infestation status. 
See Table~\ref{tab:info} and SI section S6 for more detail.

(5) \textit{Evaluate median and lower-quartile performance across counties for a range of budgets.} For all network metrics and information levels we consider a range of \textit{relative budgets}, expressed as a proportion of the maximum budget adequate to inspect all potentially infective boats in the county. 
Our unit of observation is a county. 
To measure overall performance we examine the median (0.5 quantile) and lower quartile (0.25 quantile) outcomes.
We also computed average performance across counties within the lowest quartile, and the proportion of counties where methods performed optimally or failed (defined as relative performance $< 0.66$; see Methods and SI Section S8). 

(6) \textit{Assess performance as a function of network size.}  Our last step exploits the variation in network size across counties, measured as the number of lakes within a county. Specifically, we use the results from the prior steps to examine whether performance changes with an increase in network size.

%%%%%%%%%%%%%
%% REAL RESULTS %%
%%%%%%%%%%%%%
\section*{Results}
Across all metrics and information levels metric-guided strategies achieved median performance of at least 0.80 of the optimal for reasonable budgets (relative budgets $< 0.25$; SI Table S3).
Metric-guided strategies also vastly exceeded the random benchmark, which achieved much lower performance relative to the optimal (median 0.15, lower quartile 0.10) for the same budgets (SI Table S5). As resource constraints in invasive species management mean budgets are almost never sufficient to control every site \cite{buyuktahtakin_review_2018}, we focus on results with a low relative budget (0.1 the amount required to inspect all infested boats); however, the qualitative patterns described below held for all relative budgets below 0.25 (see SI Tables S2, S3). 
Performance of metric-guided management generally improved with higher budgets (see SI Figs.~S2, S3).
Optimal management also improved with higher budgets but showed diminishing returns (SI Fig.~S1). Synthesizing the results across metrics, information levels, and network size leads to four main findings, which we describe in the following subsections.

\subsection*{Network-guided management can achieve near-optimal performance}
With full information, degree and hub+authority (H+A) achieved perfect (100\%) median performance for reasonable budgets.
Degree achieved near-perfect performance across all counties:
eomparing inspected lakes between degree and the optimal solution reveals that, for relative budgets below 0.25, degree selected the same inspection patterns as the true optimum  87\% of the time (Fig.~S4, Table S3).
A ``recalculated'' degree strategy selected the exact same inspection patterns for all relative budgets below 0.5 (see SI Section S5).
Differences between degree and H+A emerge at the lower quartile with performance of degree (100\%) exceeding H+A (97\%). Still, H+A performs well across all counties with the average performance of the lower quartile of counties 89\% of optimal.

\subsection*{Performance varies by metric}
Betweenness performed worse than degree and H+A. Across the counties, its median and lower quartile performance for a realistic budget of 0.1 of the amount required to inspect all infested boats was 93\% and 84\% of optimal, respectively (SI Table S3). Average performance across counties in the lower quartile was 72\% of optimal. Other metrics explored in the SI perform even worse, particularly in the lower quartile, revealing that performance can be tied to the choice of metric.

\subsection*{Performance can be high with less information, but it is more variable}
Metric performance 
was reduced when less detailed social and ecological information was used in the metric calculation (Fig.~\ref{fig:main}), but these patterns differed depending on which type of data was removed.
In the case with full social information but low ecological information, the performance of degree (median 97\%, lower quartile 90\%) and H+A (median 95\%, lower quartile 85\%) strategies were both high.
The performance of betweenness was lower (median 72\%; lower quartile 52\%) (SI Table S3).
In comparison, with low social but full ecological information the ranking of metrics by their median and lower quartile performance shifted. Betweenness (median 93\%; lower quartile 84\%) and H+A (median 92\%; lower quartile 88\%) performed best, with degree slightly lower (median 88\%; lower quartile 72\%). Also notable is that the average of the counties in the lower quartile was similar for H+A in each of these partial information scenarios (67\% with low environmental; 77\% with low social), and failure rates remained below 10\%. In contrast, for betweenness and degree, average performance in the lower quartile dropped to near 50\% or lower in one scenario or another, and failure rates were higher (SI Table S3).
For low social/low ecological, all focal metrics performed approximately the same, with median performance above 80\%, lower quartile approximately 65\%, and average across the lower quartile counties 44-53\% for reasonable budgets (Fig.~\ref{fig:main}A; relative budgets $< 0.25$ Table S3). 
Very similar results to the low social/low ecological case were observed 
even with only undirected topology (``minimal'' information; see SI Section S6). 

\subsection*{Performance generally does not decrease in larger-scale networks}
\begin{figure}%[\sidecaptionrelwidth][t]
\centering
\includegraphics{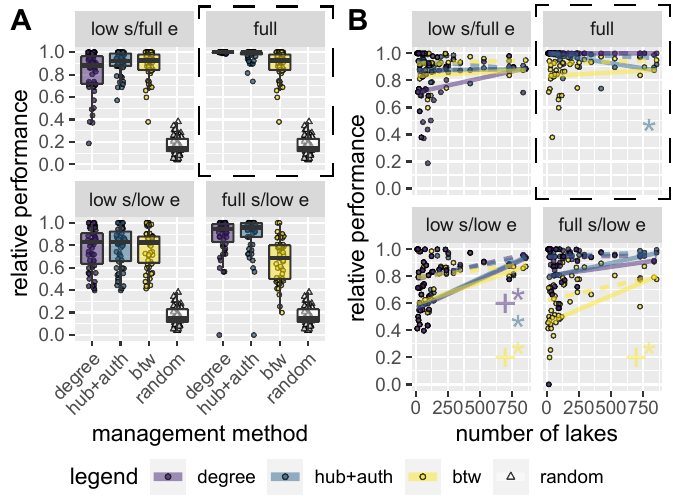}
    \caption{
Relative performance of network-guided management --measured as the number of infective boats inspected using metric-guided management as a proportion of those inspected in the integer-programming solution-- for relative budget 0.1.
Performance is shown for full information (dashed outline) and three cases with only partial information (i.e., lower levels of social and ecological information).
The legend colors apply to both panels.
(A) The distribution of performance across counties ($n=58$) illustrated with boxplots (center line, median; box limits, upper and lower quartiles; whiskers, 1.5x interquartile range) overplotted on county values (circular points). Random inspection also is included for comparison; there, triangular points represent mean performance across many random inspection placements (See Methods). See Table S2 for numeric values of median performance (0.5 quantile of the full performance distribution) and mean lower quartile performance.
(B) Relationship between performance and network size across 58 counties (circular points);
lines shown are quantile regression results for the 0.50 and 0.25 quantiles. Random inspection is not included for clarity.
{\bf *} indicates significant regression slope of the 0.25 quantile (solid line) and {\bf +} significant regression slope of the 0.50 quantile (dashed line) (no overlap of 95\% confidence interval with 0). 
    \vspace{18pt}
}\label{fig:main}
\end{figure}

Metric performance generally improves or shows no significant change as network size, measured as the number of lakes, increases (Fig \ref{fig:main}B; See Fig.~S6 for other budgets).   
Increases in performance with network size are common at the lowest information levels. In the low social/low ecological case, the lower (0.25 quantile) performance for all focal metrics was significantly higher ($p < 0.05$) for counties with larger networks (Fig \ref{fig:main}B; See Fig.~S6 for other budgets).
This finding also holds for even less information, when only the undirected topology is known (``minimal'' information; see SI Section S6, Fig.~S5). 
The one case where network size showed slight, but negative, association with performance is with full information for H+A (statistically significant decline in 0.25 quantile performance). 
However, the magnitude of this change over the range of network size was small compared to the larger positive association between performance and network size seen with low levels of ecological information. 

%%%%%%%%%%%%%%%%
%% DISCUSSION %%
%%%%%%%%%%%%%%%%
\section*{Discussion}

Our study demonstrates that network-based approaches can guide nearly-optimal invasive species management, including in large systems with constrained resources for management. We find that by managing based on some network metrics, a manager can achieve performance close to or equivalent to optimal---particularly using degree and hub + authority (H+A) (defined in Fig.~\ref{fig:overview}, results in Fig.~\ref{fig:main}A). Performance was more variable in cases with relatively low budgets and low levels of social or ecological information. In these cases, there was relatively strong median performance across metrics but greater differences in performance of counties in the lower quartiles across metrics, including cases where average performance of the counties in the lower quartile was lower than 50\% of optimal. On one hand, our results provide empirical evidence that network-guided management can achieve good performance even with limited information, but the lower-quartile results suggest that appropriate metric selection is important for achieving these outcomes.
Finally, we find that the performance of metric-guided management strategies generally does not decrease with network size and generally increases with network size with low levels of information -- i.e., in the contexts where identification of optimal solutions can be most computationally challenging.  Stability of performance across network size also suggests that metrics can be tested for performance on smaller scale systems to provide insights into preferred metrics to employ in larger systems.
Together, these results suggest that network-guided management is a promising approach for managing large-scale invasions, including those characterized by limited budgets or less detailed information about the social-ecological system. 

Our research adds to a growing body of work applying network science to the management of complex socio-environmental systems. The utility of centrality metrics for guiding invasion management, as found in our study, is expected from longstanding network theory (e.g., \cite{pastor-satorras_immunization_2002}), but their potential to guide management has not been realized for invasive species management at large scales. For small and newly-invaded systems, Chadès {\it et al.} \cite{chades_general_2011} found that centrality metrics can actually mislead management. However, large-scale invasions in real management contexts are inherently a large-network problem. While prior studies on invasive species in large networks assessed how well network metrics predict spread (e.g., \cite{banks_role_2015, perry_using_2017}) and examined performance of metrics in prioritizing management actions (e.g., \cite{kvistad2019network}), they did not rigorously compare the performance of network-based management to known-optimal interventions as we do here. 
In another conservation context (conserving species in a food web), at least one prior study compared the performance of network metric-guided management to an optimal approach \cite{mcdonald-madden_using_2016}, but did not test the performance for invasion management and under varying information, budgets, and network size, as in this study. 

Here we test network-guided management's efficacy for the objective of minimizing the number of uninspected but potentially infective boats (i.e. boats moving from infested to uninfested lakes), but other management problems could be explored in future work. Our objective captures the current means objective in our study region and is equivalent to minimizing one-period spread, which is implicitly aimed at achieving a more fundamental goal of reducing long-term damages from invasion spread \cite{MNDNR2020}. Invasive species management involves many approaches to reducing long-term impacts, with diverse management activities including surveillance, containment, and removal, and with goals spanning prevention to eradication \cite{buyuktahtakin_review_2018}. Each option represents a distinct resource allocation problem, involving different time horizons, tools, and management objectives.
Assessing the performance of network metrics for an objective like long-term damages would require integrated, dynamic socio-environmental systems models that describe changes in boater movement over time. A fuller picture of performance could also include empirical ex-post evaluation of management efficacy.

The relative performance of different metrics for network-guided management likely depends on the management objective.  
Here, the higher-performing metrics (i.e., degree and H+A in Fig.~\ref{fig:main}A) select nodes with a high weighted degree, which corresponds directly to high numbers of inspected boats.  In contrast, the lower-performing betweenness centrality depends on all paths in the network and therefore selects nodes that may not have a high degree and thus may not correspond to a high number of inspected boats. Work on the network protection problem offers further support to this idea that the best performing metric will depend on the objective. For example, when controlling spread on a network via protecting nodes \cite{nowzari_analysis_2016}, but with the objective of minimizing the rate of spread in the long-term, betweenness centrality outperforms strategies based on degree centrality \cite{holme_attack_2002}. 
Thus, our empirical results highlight the potential importance of selecting metrics that ``match'' the management objective and support a broader literature exploring this idea. 
For future research on network-guided management of invasive species or other applications of network-guided conservation, leveraging and translating predictions from network epidemiology (reviewed in \cite{nowzari_analysis_2016}) is a promising approach. 

Another outstanding topic for future research is how the performance of metric-based heuristics depends on interactions between invasive species’ dispersal characteristics and the data used to construct the network. The ordering of metric performance in cases with low levels of network information may be especially sensitive to these features. In our application, the spectral H+A metric had the most consistent performance across varying information levels. This indicates that in the partial information cases we considered (i.e., low social/full ecological or full social/low ecological; Fig.~\ref{fig:main}) H+A captures details in the network structure that result in a high ranking for nodes that turn out to be the highest degree in the full information case---i.e., on the weighted infested subnetwork. 
Future work could investigate if this finding is specific to our context, based on the network definition (from data collected over 2014-2017; \cite{kao2020}) and definition of invasion state (invaded in 2019) relative to the ecological properties (e.g., temporal and spatial dispersal abilities) of this invasive species. 
In our application, centrality metrics on the network are predictive of invaded status \cite{kao2021paper}. While this has been observed for AIS on other human transport networks \cite{kvistad2019network}, it may not be the case for some invasions.
Whether our findings extend to more general cases should be investigated, potentially through simulation approaches from network epidemiology (e.g., \cite{nowzari_analysis_2016}). 
Another promising extension would be applying our methods to networks derived from other sources. For example, gravity models have often been applied to AIS (e.g., \cite{bossenbroek2001prediction,leung_boats_2006}), and these models 
could be used to create an approximate network \cite{muirhead_development_2005} that could then be used for network-guided management.

Network approximations may be particularly valuable for spatial prioritization in very large systems, including for providing decision support for management, for several reasons. For one, integer-linear programming (ILP) approaches that have been successfully applied to small- and medium-sized problems in conservation planning (e.g., conservation prioritization with MARXAN \cite{beger2015integrating}, sparing-sharing in tropical forests with ILP \cite{runting2019larger}) have computational and data constraints that limit their use for large-scale systems. Second, complementary work by Kinsley \textit{et al.} \cite{kinseyinrev} demonstrates the effectiveness of network-guided management for real-world decision support by embedding near-optimal prioritization based on a network metric in a web-tool developed with intensive stakeholder engagement. Finally, output from network approaches such as ours could be integrated with broader frameworks for assessing invasion risk based on factors such as habitat suitability, stream connections, and impacts \cite{vander2008management,kanankege2018probability}.  Ongoing engagement between managers and researchers, such as in MN, provides a promising context for pursuing such extensions \cite{kanankege2020lessons}.

In summary, our results suggest that network-guided management could provide an important tool for addressing management challenges posed by widespread invasions, which are becoming common due to globalization and other global changes. Our study demonstrates the potential value of methods developed in other areas of network science with a richer history of working in large complex systems for invasion management. These approaches also have relevance for other environmental contexts, such as fire, fisheries, or endangered species management, where management occurs within complex systems that test the limits of traditional optimal control tools (e.g., \cite{kroetz2015bioeconomics}).
Overall our results show network approaches hold promise for improving management outcomes in contexts where modeling and/or data resources are limited.

%%%%%%%%%
% Methods
%%%%%%%%%
\newpage
\section*{Methods}

\subsection*{Data and network representation}
We use a network generated from $\approx 1.6$ million reported boater movements in Minnesota over the period 2014 to 2017 and zebra mussel lake infestation status from 2019 for 9,182 Minnesota lakes \cite{kao2020,kao2021paper}. 
This network consists of lakes (nodes) connected by directed movements of boaters (edges).
Edge weights represent estimated numbers of boaters moving between two connected lakes \cite{kao2020,kao2021paper}.
Based on lake infestation status, we categorize all network edges as either \emph{potentially infective} (i.e., a movement from an infested body of water to an uninfested body of water based on observed infestation status) or non-infective (see SI Sections S1 \& S2).

\subsection*{
Defining county-level networks and infested subnetworks
}
We construct 58 county-level networks from the Minnesota-wide data, including all counties with greater than 10 lakes. We account for out-of-county and in-to-county boater movement by adding two nodes to each county network that serve as synthetic lakes: the first corresponding to all non-county lakes that are uninfested and the second corresponding to all non-county lakes that are infested. We aggregate all edge-weight that crosses a county line into the incoming and outgoing links to these two synthetic nodes. 
In this way, we define a county-level network as an adjacency matrix \(A^{(k)}\) (for \(k\) in 58, the number of counties that we examine). These county-level networks vary widely in summary statistics related to the network and infestation status, including number of edges, number of trips, and proportion of potentially infective trips (see Table S1).

For each county, we further define an \emph{infested subnetwork}, which is a subgraph of each full county network \(A^{(k)}\), consisting of only the potentially infective edges and their connected lakes. Each edge in this subnetwork is a directed edge from an infested lake to an uninfested lake. The subnetwork connects a subset of lakes in the county network as an adjacency matrix \(\widetilde{A}^{(k)}\). The entry \({\widetilde{A}}^{(k)}_{\text{ij}}\) corresponds to the number of trips from lake \(i\) to lake \(j\).

\subsection*{Problem formulation}
Although boater movement connects lakes across counties, resource allocation decisions primarily occur within counties, so we study a decision problem at that scale. The decision problem for county \(k\) is to determine the location of inspection stations, given a limited budget \(B^{(k)}\), to maximize the number of boats inspected that are moving from infested lakes anywhere to uninfested lakes within the county.
We assume uniform costs to inspect any given lake (i.e., the cost of inspection stations are the same for each lake).

\subsection*{Optimal solution and random-inspections benchmark} 
As a reference point and the upper bound on the performance of our network metrics we solve for the optimal solution for a fixed budget as an integer linear program, an approach that has previously been applied to optimize inspection for invasive species \cite{fischer2021managing,haight2021paper}.
Indeed, the ILP formulation for the optimal solution that we use here was developed to support AIS management in several MN counties (\cite{haight2021paper}; also see SI Section S3).
For each county $k$, we use the adjacency matrix of the infested subnetwork \({\widetilde{A}}^{(k)}\) to define the \emph{infested matrix} \(N^{(k)}\) with rows corresponding to the infested lakes, within the county and the infested synthetic lake, and columns corresponding to the uninfested lakes only in the county.
Each entry \(N_{\text{ij}}\) gives the number of boats moving from infested lake \(i\) to uninfested lake \(j\).
We define the decision problem of locating inspection stations as a sum across entries in this matrix, subject to constraints for the budget and to disallow inspection of synthetic lakes as an integer-linear program (see SI Section S3).

As a lower bound on performance, we also computed the performance of randomly-located inspection stations, where for a budget of $B^{(k)}$, we chose $B^{(k)}$ lakes at random from all lakes in county $k$. To determine the mean performance of a random strategy for a given budget, we initially compute the mean performance for 100 replicates, then add replicates in increments of 5 until mean performance across all replicates changes by less than one inspected boat. Plots for the random strategy show the final mean performance; for most county-by-budget combinations, this procedure resulted in a final mean taken over 105 to 570 replicates.

\subsection*{Network metric solution}
We compute metric guided strategies degree, hub+auth, and betweenness (Table \ref{tab:metrics}) as described in the SI Sections S4 \& S5). 
For each metric and each information level, we select nodes for inspection sequentially until the budget (total number of nodes that can be inspected) is reached. 
For all metrics we select nodes based on the ranking from the initial state of the network (see SI Algorithm 1).
We also examined a ``recalculated'' method \cite{holme_attack_2002}:
compute the metric for all nodes, select the node with the highest metric score for inspection, update the network to exclude the selected node, recalculate the metric for all remaining uninspected nodes, then select the remaining node that has the highest score, continuing with this process until the budget is reached (see SI Algorithm 2).
For the objective and metrics we examine here, the ``recalculated" method shows little improvement over the initial state method (see SI Section S5, Fig.~S7).

\subsection*{Computing relative performance of non-optimal strategies}
To compare the performance of network-guided management (i.e., using heuristics based on centrality metrics) to the optimal allocation, we compute the number of potentially infective  boats that are inspected under each strategy. We calculate relative performance as the number of infective boats inspected using metric-guided management divided by the number of infective boats inspected using the optimal integer-programming approach (see SI S7).
We apply this approach for each county $k$ and for each budget up to the minimum budget required to inspect all potentially infective boats given full information and solving the inspection problem for the optimal solution. We denote this budget $B^{(k)}_\text{cover}$ (See SI Section S3).
We use the maximum budget to define each of the $i$ county's relative budget $b^{(k)} = B^{(k)}/B^{(k)}_{\text{cover}}$, which facilitates comparing amongst counties of different sizes.
We evaluated performance by calculating the median (0.5 quantile) relative performance across all counties for each budget as well as mean lower quartile (0.25 quantile) performance.
    We also computed the proportion of counties where the performance is perfect, i.e., matching the optimal exactly, and the proportion of failures (defined as relative performance $< 0.66$; see SI Section S8).
    These values are reported in SI Tables S2--S5.

\subsection*{Quantifying how performance depends on network characteristics}
Quantile regression estimates relationships between predictors and parts of the outcome distribution other than the mean \cite{cade2003gentle}.
To test associations between relative performance both at the median (0.5 quantile) and lower quartile (0.25 quantile) and network characteristics (size or the number of nodes; average degree) we computed quantile regressions, using R package {\tt quantreg}. 
We performed linear quantile regressions for the median and lower quartile (quantiles 0.5, 0.25) and tested for significance using confidence intervals produced via inverting a rank test \cite{koenker1994confidence}. 
This method quantifies whether network characteristics show a significant association with increases or decreases in both average and lower-end performance.

\section*{Data availability}
The network data used in this study were previously reported and are available in \cite{kao2020}.
The specific data supporting this study, including network data, lake metadata including infestation status, and geospatial data delineating county boundaries are available in figshare with identifier \href{https://figshare.com/s/9bfd4183e41810412596}{10.6084/m9.figshare.14402447}.
% note private link must be swapped for publication

\section*{Code availability}
Analysis employed R version 4.0.2 (2020-06-22) using packages  dplyr (v1.0.7), purrr (v0.3.4), ggplot2 (v3.3.3), igraph (v1.2.5), quantreg (v5.61), Rglpk (v0.6-4).
Full analysis code underlying all analyses is available in figshare with
identifier \href{https://figshare.com/s/9bfd4183e41810412596}{10.6084/m9.figshare.14402447}.
% note private link must be swapped for publication

\section*{Correspondence and requests for materials}

All correspondence and requests for materials should be addressed to Jaime Ashander.

\section*{Acknowledgements}

The authors thank A. Kinsley for comments on a previous draft.
Funding for this research was provided by Resources for the Future and the National Socio-Environmental Synthesis Center (SESYNC) under funding received from the National Science Foundation (NSF) DBI-1639145.
The Northern Research Station, USDA Forest Service also provided support. L.E.D. acknowledges support from NSF OCE-2049360.

\section*{Contributions}

J.A., L.E.D, K.K.  conceived the study; 
J.A., L.E.D, R.E.-N., K.K. designed research;  
N.B.D.P., R.G.H. contributed data or analytic tools; 
J.A. performed research;
J.A., L.E.D, R.E.-N., K.K., wrote the paper; and 
all authors edited the paper.

\section*{Tables}

\begin{table}
\caption{\textbf{Network centrality metrics.}
Definitions for network centrality metrics used in our three focal strategies, which
prioritize nodes with highest centrality value.
}
\label{tab:metrics}
\centering
\begin{tabular}[t]{ p{5cm}p{10cm}}
\toprule
%Budget category & Metric & Median & Lower quartile & Average $<25$\% & Proportion perfect & Proportion $<0.66$\\
Method &Description\\
\midrule
Degree      &\textit{Degree} is the number of direct links (incoming and out-going) incident on a
node. In weighted networks, the equivalent is \textit{weighted degree} or \textit{strength}.\\
\hline \\
Hub Score + Authority \newline Score (H+A)  &Sum of two eigenvalue-based metrics that account for incoming links
(\textit{Authority Score}) and out-going links (\textit{Hub Score}). Unlike degree, the
contribution of a direct link is weighted by the linking node’s score. \\
\hline \\
Betweenness (btw)   & The number of shortest paths between all other nodes that pass through a
node.\\
\bottomrule
\end{tabular}
\end{table}

\begin{table}
\caption{\textbf{Varying social and ecological information.}
Our four focal cases of full or partial information result from varying the level of detail on ecological and social information between two levels defined in this table.
(See also Fig \ref{fig:overview}C and SI section S6).
}
\label{tab:info}
\centering
\begin{tabular}[t]{ p{3cm}p{8cm}}%|p{1cm}|p{1.25cm}|p{1.25cm}|p{1.25cm}}
\toprule
Information \newline level   &Definition \\
%                             &\textbf{full}
%                             &\textbf{low s/ \newline full e} 
%                             &\textbf{full s/ \newline low e}
%                             &\textbf{low s/ \newline low e}\\
\midrule
Full social     &The weighted-directed network of boat movements (edges represent the number of boats moving from source to target lake).\\
%&X & &X &\\
\hline \\
Low social      &Unweighted-directed network of boat movements (edges represent one or more boat moving from source to target lake).\\
%& &X & &X\\
\hline \\
Full ecological &Lake position in network and invasion status.\\
%&X &X & &\\
    \hline \\
Low social      &Lake position in network only.\\
%& & &X &X\\
\bottomrule
\end{tabular}
\end{table}

\clearpage
\appendix

\renewcommand{\thesection}{S\arabic{section}}
\setcounter{figure}{0}    
\setcounter{table}{0}    
\renewcommand\thetable{S\arabic{table}}
\renewcommand\thefigure{S\arabic{figure}}

\section{Background on the Zebra Mussel Invasion and Management Context in Minnesota, USA}\label{si:zm}
Zebra mussel invasions occurring at the landscape scale provide an ideal application to test the utility of our framework. Since their initial introduction into the United States (U.S.) in the mid-1980s from contaminated ballast water discharge into the Great Lakes, zebra mussels have spread rapidly to at least 409 sub-watersheds throughout the country, facilitated by natural (e.g., water connections \cite{bobeldyk_secondary_2005}) and human-mediated pathways (e.g., boater movements \cite{kanankege2018probability, mallez2018dispersal}). 
The latter is the dominant form of dispersal \cite{kanankege2018probability, mallez2018dispersal} and is particularly concerning given the long-distance dispersal potential of the microscopic juvenile stage in residual boat water and the hardy adult stage that attaches to hard surfaces (e.g. boat hulls) (\cite{johnson_overland_2001}).  Following patterns throughout North America, the lake-river systems of Minnesota (MN), USA, have become infested with zebra mussels, with \(\approx 50\)\% of the infestations occurring since 2016 \cite{MNDNR2020}. Representative of the national patterns, this rapid expansion and overland spread is closely associated with movements of recreational boats (as recreational boating is a major activity and source of tourism revenue in MN) \cite{kanankege2018probability}. 
    
Once introduced and established, zebra mussels often become highly abundant, causing changes to ecosystem processes \cite{mceachran2019stable}, extirpating native mussel species \cite{karatayev2002impacts}, and damaging infrastructure \cite{prescott2013impact}. Although poorly quantified, this invasive species costs managers many millions of dollars each year \cite{chakraborti2016costs}. While small-scale eradications have been attempted \cite{lund2018zebra}, there are currently no effective removal methods for established populations. Therefore, in Minnesota and many other places, the primary management objective is preventing the introduction of zebra mussel to new waterbodies \cite{leung_ounce_2002,bobeldyk_secondary_2005}. Cost-effective management strategies are urgently needed. 

 Currently, significant funding from the state is allocated at the county and local levels for intervention: \$10M annually is divided among 87 counties for management. To date, a significant portion of zebra mussel management in MN is conducted by county planning organizations that each receive a funding allocation from the state, which is generally spent on boat inspection programs. In these inspection programs, trained officers are authorized to inspect a watercraft and related equipment at a water-access site, any other public location in the state, or in plain view on private property (2018 Minnesota Statutes 84D.105). Individuals transporting zebra mussels are subject to a citation, including a fine depending on the offense (Minnesota Rules, 2018; MN § 84D.13).
 
 Once introduced and established, zebra mussels often become highly abundant, re-engineer ecosystems \cite{mceachran2019stable}, extirpate native mussel species \cite{karatayev2002impacts}, and significantly impact industrial and recreational infrastructure \cite{prescott2013impact}. In MN, zebra mussel is a prohibited invasive species occupying around 3\% of waters (over 230 lakes/rivers). Putting many other waterways at risk, zebra mussel spread is primarily facilitated by boater movements \cite{kanankege2018probability, mallez2018dispersal}. Thus MN’s specific management objective is to contain the invasion by maximizing the number of potentially-infective boats that are inspected and cleaned, for a fixed budget. The budget is allocated to establishing and staffing boat inspection stations, for a fixed cost, at selected lakes. These inspect and clean all incoming or departing boats to prevent boats from moving zebra mussels from infected to uninfected lakes. The cost of inspection units is uniform as the fixed cost of setting up a station and staffing varies little over space and one staffer can generally handle the necessary inspections in any given location.
Here we investigate how to allocate the substantial, but insufficient funding, available at local levels for intervention (\$10M annually divided among 87 counties). 
This application is ideal for our aims of assessing network-guided management approaches, as the counties in MN differ in system size (number of nodes and connectivity) and the extent of the invasion.

\section{Data on Minnesota boat movement networks and lake infection status}\label{si:data}
Our dataset is based on previously estimated boater movements between lakes and infection status data for 9,182 Minnesota lakes \cite{kao2020}. To define the networks of boater movement between the Minnesota lakes used in the analysis, boater movements were estimated by two-stage boosted regression trees \cite{kao2021paper}. The movement estimate ---the estimated number of boats moving between each lake--- for the statewide network is output from the boosted regression procedure as an edge list with weight corresponding to the point estimate of the number of boat movements between two waterbodies.
We define a lake's infection attribute, using data on Zebra Mussel infection for all Minnesota lakes in 2019 \cite{kao2020}, as infected or not. Based on lake infection status, we further categorized all edges as either \emph{potentially infecting} (i.e., a movement from an infected body of water to an uninfected body of water based on observed infection status) or non-infecting. 
We treat movement of recreational boats as a weighted, directed network \((\mathcal V, \mathcal E)\) with lakes as nodes \(\mathcal V\) and boat movement as edges \(\mathcal E\); the edge weight, \(W(e)\) for \(e \in \mathcal E\), defines the number of boat movements along any edge. A lake's infection status is encoded as a binary node attribute \(\mathcal I(v)\) for \(v \in \mathcal V\).

\section{Optimal inspection problem as an integer-linear program}\label{si:problem}

We consider the 
\emph{infective matrix} \(N^{(k)}\) with rows corresponding to the infected lakes, both within the county and the infected synthetic lake, and columns corresponding to the uninfected lakes only in the $k$-th county.
Each entry \(n_{\text{ij}}\) gives the number of boats moving from infected lake \(i\) to uninfected lake \(j\).
To define the decision problem, let \(x_{i}\) be a binary choice variable denoting inspection (\(x_{i} = 1\)) or not (\(x_{i} = 0\)) of infested lake \(i\), and \(y_{j}\) be a binary decision variable denoting inspection of uninfested lake \(j\). Then we also introduce binary variable \(u_{\text{ij}}\) that denotes if a given edge is
inspected by inspections at either the associated infested or uninfested lake or both, as indicated by the decision variables \emph{x\textsubscript{i}} and \emph{y\textsubscript{j}}. Further we denote the number of columns as \(J\) and the number of rows as \(I\)
and assume that the \(I\)-th row of the matrix \(N\) includes all boats originating from infected lakes outside of the county. Then the county decision problem is to choose \(x_{i},y_{j},u_{\text{ij}} \in \{ 0,1\}\)
to maximize the number of inspections of boats moving from infested to uninfested lakes, 

\begin{align*}
&\max_{x,y,u} \sum_{i,j} n_{ij}u_{ij}\\
      &s.t.: \sum_{i,j} x_i + y_j \leq B^{(k)} \\
      &u_{ij} \leq x_i + y_j \; \forall \; i \in I \; and \; j \; \in \; J \numberthis \label{opt} \\
       &x_I = 0 
\end{align*}

\noindent when the first constraint imposes a fixed maximum budget $B^{(k)}$ and each inspection station costing one unit of budget, the second set of I*J constraints ensures that infected boats traveling along an edge will be counted as inspected if the origin or destination node (lake) is chosen for inspection, and the last constraint ensures that no inspection station can be located at the synthetic lake.

\subsection*{Statement in matrix form}\

\textbf{Choice variables}: To restate the problem in standard (or canonical) matrix form, we pose the vector of choice variables $X = (u_{\text{ij}}|y_{j}|x_{i})$ where the $u_{\text{ij}}$ are ordered by index $(i - 1)*I + j$ column-major order.

\textbf{Objective}: the objective is \(a^{T}X\) where $a \in R^{2IJ + I + J}$ with elements $a = (n_{\text{ij}}|0_{j}|0_{i}),$ with \(n_{\text{ij}}\) ordered as column-major.

\textbf{Constraints}: The constraint equation is of matrix form with a
matrix of constraint coefficients $C \in R^{(IJ + 1 + 1) \times (IJ + I + J)}$ with right-hand-side $b \in R^{IJ + 1 + 1}$: $\text{CX} =^{?}b$.

The notation `\(=^{?}\)' reflects that the type of constraint can differ by row. Within \(C\):
\begin{itemize}
\item The first \(\text{IJ}\) rows encode the constraints \textgreater{}
\(u_{\text{ij}} - y_{j} - x_{i} \leq 0\),
\item The next row encodes the budget \(x + y \leq B\)
\item The final row encodes the synthetic lake constraint \(y_{J} = 0\)
\end{itemize}

\subsection*{Covering Budget and Relative Budget}
To facilitate comparisons among counties for different budgets, we define a relative budget for county $k$, \(b^{(k)} = B^{(k)}/B^{(k)}_{\text{cover}}\) where \(B^{(k)}_{\text{cover}}\) is the budget that is adequate to inspect all potentially infective boat movements for county $k$. To compute \(B^{(k)}_{\text{cover}}\) we solve a related decision problem, also using integer-programming, with the objective to minimize the cost of inspecting all infective boat movements. Using the same variables in \eqref{opt}, the covering budget \(B_{\text{cover}}\) problem is formulated as:
\begin{align*}
&\min_{x,y,u} \sum_{i,j} x_i +y_j\\
      &s.t.: u_{\text{ij}} \geq 1 \; \forall \; i \in I-1 \;\text{and} \; j \in J \\
      &u_{ij} \leq x_i + y_j \forall \; i \in I \; and \; j \; \in \; J \numberthis \label{budgetcover} \\
       &x_I = 0 
\end{align*}
\noindent where the first (I-1)*(J) constraints require all infective edges be inspected at least once, the second set of I*J constraints ensures that infected boats traveling along an edge will be counted as inspected if the origin or destination node (lake) is chosen for inspection, and the last constraint ensures that no inspection can occur at the synthetic node. Results for relative budget of 0.1 are shown in the main text. 

\section{Network metrics}\label{si:metrics}
Centrality metrics aim to quantify the importance of a given node to the structure and/or dynamics on a network. The precise interpretation of ``importance'' varies depending on the metric. Although metrics can be computed on any static network, many metrics have a dynamic interpretation (e.g., involving potential paths or random walks). As noted in the main text, such metrics are naturally attractive as heuristics for managing spreading phenomena on networks including invasive species \cite{albert2000error}, however the problem of selecting a maximally influential set of nodes is much harder than finding the most single influential node \cite{pei_theories_2018}. 

We compute network metrics for each county network, for each of four cases that have different levels of information (Fig. 1).
In each case we use \texttt{igraph} to compute the metrics (\cite{igraph2006}). The metrics we use include several well-known centrality metrics that aim to quantify the structural and dynamic importance of nodes. The specific metrics are as defined as follows:

\begin{itemize}
    \item \emph{Degree}: The node total weighted degree (i.e., total strength), is the sum of the weights of incoming and outgoing links.
    \item \emph{Betweenness}: The node betweenness which is roughly the number of shortest paths incident on a node. Weights, if present, are interpreted as the inverse path length between two nodes (\texttt{igraph}; \cite{freeman1978centrality}) 
    \item \emph{PageRank}: The node's PageRank is its entry in the dominant eigenvector of the normalized adjacency matrix \(A\) (normalized by the node outdegree). The actual computation is via an iterative algorithm (\texttt{igraph}; \cite{brin1998anatomy}). This is a spectral metric.
\end{itemize}

We also propose and apply two novel spectral metrics that combine the Hub and Authority Scores of Kleinberg in \cite{kleinberg1998authoritative}. The hub+authority score (H+A) favors nodes that are {\it either} sources of infection or targets of infection on a directed network. The hub$\times$authority score (H$\times$A) favors nodes that are {\it both } sources of infection and targets of infection on a directed network.
Like eigenvector centrality these are both spectral metrics and so recursively weight a node's importance by the importance of its linked nodes. 
These novel metrics are as follows:

\begin{itemize}
    \item \emph{Hub + Authority}: This is the sum of the hub score and authority score, which are based on the eigenvectors of the adjacency matrix \(A\) of the graph.
    The \emph{hub} score is the node's entry in the dominant eigenvector of \(AA^{T}\). The \emph{authority} score is the node's entry in the dominant eigenvector of \(A^{T}A\). Note for undirected graphs (i.e.~symmetric \(A\)) the hub score is equivalent to the \emph{eigenvector} centrality up to a scaling factor (\texttt{igraph}; \cite{kleinberg1998authoritative}; extended discussion in \cite{kleinberg1999authoritative}). 
    \item \emph{Hub $\times$ Authority}: This is the product of hub and authority. In contrast to hub+auth it favors only nodes that have {\em both} hub and authority properties. 
\end{itemize}

\section{Heuristic network-metric-based strategies for management}\label{si:heuristics}
For comparison to the optimal solution, we considered heuristic solutions based on each of the network metrics (see Network Metrics, above). 

\subsection*{The initial state method presented in the main text}
For all metrics including degree, betweenness, and spectral metrics (e.g., Page Rank, hub+auth, hub*auth), we applied heuristic inspection strategy based on the initial state of the network, i.e., the centrality values before any inspection decisions.

For a network-metric based solution based on the \emph{initial state} of the network, a metric for each node is calculated from the network structure prior to any inspection location placement and those values are used throughout the solution procedure. This approach is defined in Algorithm \ref{alg:initial} for a fixed budget $B$. 
The initial state inspection method is used for the all metrics presented in the main text. It is also used for the hub $\times$ authority (H$\times$A) and PageRank strategies presented in the SI.

\begin{algorithm}[H] % required for revtex
\caption{Initial state inspection based on metric $f$ applied to network $N$ for budget $B$.}\label{si:alg:initial}
\begin{algorithmic}
\Procedure{Metric Rank}{$N, f$}\Comment{Determine the inspected nodes.}
\State $M \gets f(N)$ \Comment{Compute vector of node level metrics}
\State Compute $\mathbf k$ as a vector of node indices ordered their metric value $M$ in descending order
\State $\mathbf i$ a node attribute vector indexed by node indices \Comment{Defining inspected}
\State $\mathbf i[n] = 0$ for all nodes $n \in N$
\For{$i \in 1 \dots B$}
\State $\mathbf i[\mathbf k[i]] = 1$ for the $i$-ranked node 
\EndFor
\EndProcedure
\end{algorithmic}
\end{algorithm}

\subsection*{The recalculated method and its relative performance}

We also examined a \emph{recalculated} heuristic inspection strategy, where the metric and ranking are recalculated after each inspection placement decision.
The \emph{recalculated} inspection strategy calculates the metric and ranking after each decision and is defined in Algorithm \ref{alg:recalc} for a fixed budget \(B\).
Results from the recalculated inspection method for all metrics are presented in the SI (Fig.~\ref{si:fig:recalc}).
\begin{algorithm}[H]
    \caption{Recalculated based on metric $f$ applied to network $N$ for budget $B$.}\label{si:alg:recalc}
\begin{algorithmic}
\Procedure{Recalculated Metric Rank}{$N, f$}\Comment{Determine the inspected nodes.}
\State $M \gets f(N)$ \Comment{Compute vector of node level metrics}
\State Compute $\mathbf k$ as a vector of node indices ordered their metric value $M$ in descending order
\State $\mathbf i$ a node attribute vector indexed by node indices \Comment{Defining inspected}
\State $\mathbf i[n] = 0$ for all nodes $n \in N$
\State $N' = N$ a copy of the network for modification.
\For{$i \in 1 \dots B$}
\State $\mathbf i[\mathbf k[1]] = 1$ for the top-ranked node 
\State Remove node $N'[\mathbf k[1]]$ from $N'$
\State Remove index $\mathbf k[1]$ from $\mathbf k$
\State Compute $M'$ from $N'$ \Comment{$N'$ now modified}
\State Reorder the $\mathbf k$ by metric value $M'$ in descending order.
\EndFor
\EndProcedure
\end{algorithmic}
\end{algorithm}

We examined the relative performance of recalculated and initial state strategies. 
Recalculated (also called ``adaptive'') strategies have been shown to improve performance of degree and betweenness \cite{holme_attack_2002}, albeit for different objectives. 
Recalculated methods also improve performance of metrics for detecting influential nodes across a corpus of real world networks from social, technology, biological, and transportation domains \cite{erkol2019systematic}, but only when invasion dynamics are not favorable for spread (technically, when the network dynamics are in the sub-critical phase where the probability of spread is less than the critical level required for an invasion to spread network-wide).  
The recalculated methods improves performance of some metrics.
In fact, given the focal management objective, degree used with a recalculated method is actually identical to the greedy approximation algorithm guaranteed to give best approximate performance (to a factor of $1-\tfrac{1}{e}$) for the maximum coverage problem in our full information case \cite{feige1998threshold}.
However, on average we found little improvement (Fig.~\ref{si:fig:recalc}), which is consistent with an established invasion where dynamics are favorable for spread. Therefore, for our main results, we focus on the metrics without recalculation because they are easier to calculate.

\section{Information scenarios}\label{si:info}

We considered four cases with varying levels of information corresponding to increasing data on the social dimension (the boat movement network) and ecological dimension (infection status of each lake). The full weighted, directed network is encoded by the adjacency matrix \(A\) (suppressing the indices \(i\) for each
county). Additional information on infection status (ecological information) is needed to obtain the \emph{infested network}, encoded by \(\widetilde{A}\). 
With less social information, only the unweighted network is known; it is defined by a (binary) adjacency matrix \(B\) where \(B_{ij} = 1\) if \(A_{ij} > 0\) and otherwise \(B_{ij} = 0\)
The varying information levels we consider are as follows: 

\begin{itemize}
    \item full information: all ecological and social information used in the optimal linear programming solution; the directed and weighted topology of the network is known, and the infection status of each lake is known as well, so the invective network \(\widetilde{A}\) is known.
\item full social/low ecological: the directed and weighted topology of the network corresponding to the adjacency matrix \(A\) is known.
\item low social/full ecological: the directed, unweighted topology of the network is known, as is the infection status. So the binary infested network \(\widetilde{B}\) is known.
\item low social/low ecological: the directed, unweighted topology of the network is known, this corresponds to the binary adjacency matrix $B$.
\item minimal: only the unweighted undirected topology of the network is known. This corresponds to knowing the matrix $S$, the symmetrized binary adjacency matrix $S := \sqrt{B^TB}$ 
\end{itemize}

Results for minimal are quite similar to low social/low ecological and are only shown in the SI.

\section{Comparing the optimal and network-metric based solutions}\label{si:comparing}

Any metric-based strategy as defined above yields a network with node attribute inspection (0 for uninspected, 1 for inspected). Using this attribute, we compute the performance of the network-metric based solution on the same terms as the integer-programming solution: the total number of potentially infective boats inspected. We do this by directly using the objective function of the IP problem Equation \eqref{opt}), and assigning $u_{ij} = 1$ for any entry where lake \(i\) or lake \(j\) is inspected.

Because we compute the performance of the network-metric based solution using the objective function, comparison with the integer-programming solution on these terms is straight-forward. We focus on \emph{relative performance} measured by the objective function. Denoting the IP-optimum value of the objective function as $\text{Obj}_\text{IP}$ and the value of the objective for a given network-metric based method as $\text{Obj}_{\text{metric}}$, relative performance is \({\text{Obj}}_{\text{metric}} /\text{Obj}_{\text{IP}}\). 
To characterize degree of similarity between the inspection locations chosen by network-metric based solutions, we also computed their accuracy relative to the optimal integer-programming solution. Accuracy is computed as the proportion of matching sites (inspected or uninspected) between the metric and optimal solution (Fig \ref{si:fig:acc}).
We examine relative performance and accuracy of the metrics across all counties and for varying values of the relative budget \(B_{\text{rel}}\) up to \(B_{cover}\).

\section{Additional results on distribution of performance}\label{si:worst}

We quantified agreement with optimal, failure rate, and expected performance in the worst 25\% of counties for each metric, budget, and information scenario. 
To understand best-case performance, we examined the proportion of counties where the network metrics result in perfect performance---inspections of the same number of infective boats as the optimum ILP.
To quantify performance in the worst cases, we also examined the failure rate (defining ``failure'' as $< 0.66$ of relative performance)
as well as the mean performance in the lower quartile (mean of all counties below 0.25 quantile in the relative performance distribution), for each metric, budget, and information scenario.
See Table \ref{si:tab:results} for full results with varying budgets and Table \ref{si:tab:results-bin} for results in budget bins.
Results for random are shown in Table \ref{si:tab:results-random} and \ref{si:tab:results-random-bin}.

\bibliographystyle{elsarticle-harv} 
\bibliography{inspect}

\begin{thebibliography}{63}
\expandafter\ifx\csname natexlab\endcsname\relax\def\natexlab#1{#1}\fi
\providecommand{\url}[1]{\texttt{#1}}
\providecommand{\href}[2]{#2}
\providecommand{\path}[1]{#1}
\providecommand{\DOIprefix}{doi:}
\providecommand{\ArXivprefix}{arXiv:}
\providecommand{\URLprefix}{URL: }
\providecommand{\Pubmedprefix}{pmid:}
\providecommand{\doi}[1]{\href{http://dx.doi.org/#1}{\path{#1}}}
\providecommand{\Pubmed}[1]{\href{pmid:#1}{\path{#1}}}
\providecommand{\bibinfo}[2]{#2}
\ifx\xfnm\relax \def\xfnm[#1]{\unskip,\space#1}\fi
%Type = Article
\bibitem[{Aadland et~al.(2015)Aadland, Sims and Finnoff}]{aadland2015spatial}
\bibinfo{author}{Aadland, D.}, \bibinfo{author}{Sims, C.},
  \bibinfo{author}{Finnoff, D.}, \bibinfo{year}{2015}.
\newblock \bibinfo{title}{Spatial dynamics of optimal management in bioeconomic
  systems}.
\newblock \bibinfo{journal}{Computational Economics} \bibinfo{volume}{45},
  \bibinfo{pages}{545--577}.
%Type = Article
\bibitem[{Albert et~al.(2000)Albert, Jeong and Barab{\'a}si}]{albert2000error}
\bibinfo{author}{Albert, R.}, \bibinfo{author}{Jeong, H.},
  \bibinfo{author}{Barab{\'a}si, A.L.}, \bibinfo{year}{2000}.
\newblock \bibinfo{title}{Error and attack tolerance of complex networks}.
\newblock \bibinfo{journal}{Nature} \bibinfo{volume}{406},
  \bibinfo{pages}{378--382}.
%Type = Article
\bibitem[{Baker(2017)}]{baker2017target}
\bibinfo{author}{Baker, C.M.}, \bibinfo{year}{2017}.
\newblock \bibinfo{title}{Target the source: optimal spatiotemporal resource
  allocation for invasive species control}.
\newblock \bibinfo{journal}{Conservation Letters} \bibinfo{volume}{10},
  \bibinfo{pages}{41--48}.
%Type = Article
\bibitem[{Banks et~al.(2015)Banks, Paini, Bayliss and Hodda}]{banks_role_2015}
\bibinfo{author}{Banks, N.C.}, \bibinfo{author}{Paini, D.R.},
  \bibinfo{author}{Bayliss, K.L.}, \bibinfo{author}{Hodda, M.},
  \bibinfo{year}{2015}.
\newblock \bibinfo{title}{The role of global trade and transport network
  topology in the human-mediated dispersal of alien species}.
\newblock \bibinfo{journal}{Ecology Letters} \bibinfo{volume}{18},
  \bibinfo{pages}{188--199}.
\newblock \URLprefix \url{http://doi.wiley.com/10.1111/ele.12397},
  \DOIprefix\doi{10.1111/ele.12397}.
%Type = Article
\bibitem[{Beger et~al.(2015)Beger, McGowan, Treml, Green, White, Wolff, Klein,
  Mumby and Possingham}]{beger2015integrating}
\bibinfo{author}{Beger, M.}, \bibinfo{author}{McGowan, J.},
  \bibinfo{author}{Treml, E.A.}, \bibinfo{author}{Green, A.L.},
  \bibinfo{author}{White, A.T.}, \bibinfo{author}{Wolff, N.H.},
  \bibinfo{author}{Klein, C.J.}, \bibinfo{author}{Mumby, P.J.},
  \bibinfo{author}{Possingham, H.P.}, \bibinfo{year}{2015}.
\newblock \bibinfo{title}{Integrating regional conservation priorities for
  multiple objectives into national policy}.
\newblock \bibinfo{journal}{Nature Communications} \bibinfo{volume}{6},
  \bibinfo{pages}{1--8}.
%Type = Article
\bibitem[{Bobeldyk et~al.(2005)Bobeldyk, Bossenbroek, Evans-White, Lodge and
  Lamberti}]{bobeldyk_secondary_2005}
\bibinfo{author}{Bobeldyk, A.M.}, \bibinfo{author}{Bossenbroek, J.M.},
  \bibinfo{author}{Evans-White, M.A.}, \bibinfo{author}{Lodge, D.M.},
  \bibinfo{author}{Lamberti, G.A.}, \bibinfo{year}{2005}.
\newblock \bibinfo{title}{Secondary spread of zebra mussels (
  \textit{{Dreissena} polymorpha} ) in coupled lake-stream systems}.
\newblock \bibinfo{journal}{Écoscience} \bibinfo{volume}{12},
  \bibinfo{pages}{339--346}.
\newblock \URLprefix
  \url{https://www.tandfonline.com/doi/full/10.2980/i1195-6860-12-3-339.1},
  \DOIprefix\doi{10.2980/i1195-6860-12-3-339.1}.
%Type = Article
\bibitem[{Bodin et~al.(2019)Bodin, Alexander, Baggio, Barnes, Berardo, Cumming,
  Dee, Fischer, Fischer, Garcia et~al.}]{bodin2019improving}
\bibinfo{author}{Bodin, {\"O}.}, \bibinfo{author}{Alexander, S.M.},
  \bibinfo{author}{Baggio, J.}, \bibinfo{author}{Barnes, M.L.},
  \bibinfo{author}{Berardo, R.}, \bibinfo{author}{Cumming, G.S.},
  \bibinfo{author}{Dee, L.E.}, \bibinfo{author}{Fischer, A.},
  \bibinfo{author}{Fischer, M.}, \bibinfo{author}{Garcia, M.M.}, et~al.,
  \bibinfo{year}{2019}.
\newblock \bibinfo{title}{Improving network approaches to the study of complex
  social--ecological interdependencies}.
\newblock \bibinfo{journal}{Nature Sustainability} \bibinfo{volume}{2},
  \bibinfo{pages}{551--559}.
%Type = Article
\bibitem[{Bossenbroek et~al.(2001)Bossenbroek, Kraft and
  Nekola}]{bossenbroek2001prediction}
\bibinfo{author}{Bossenbroek, J.M.}, \bibinfo{author}{Kraft, C.E.},
  \bibinfo{author}{Nekola, J.C.}, \bibinfo{year}{2001}.
\newblock \bibinfo{title}{Prediction of long-distance dispersal using gravity
  models: zebra mussel invasion of inland lakes}.
\newblock \bibinfo{journal}{Ecological Applications} \bibinfo{volume}{11},
  \bibinfo{pages}{1778--1788}.
%Type = Article
\bibitem[{Brin and Page(1998)}]{brin1998anatomy}
\bibinfo{author}{Brin, S.}, \bibinfo{author}{Page, L.}, \bibinfo{year}{1998}.
\newblock \bibinfo{title}{The anatomy of a large-scale hypertextual web search
  engine}.
\newblock \bibinfo{journal}{Computer networks and ISDN systems}
  \bibinfo{volume}{30}, \bibinfo{pages}{107--117}.
%Type = Article
\bibitem[{Bushaj et~al.(2021)Bushaj, B{\"u}y{\"u}ktahtak{\i}n, Yemshanov and
  Haight}]{bushaj2021optimizing}
\bibinfo{author}{Bushaj, S.}, \bibinfo{author}{B{\"u}y{\"u}ktahtak{\i}n,
  {\.I}.E.}, \bibinfo{author}{Yemshanov, D.}, \bibinfo{author}{Haight, R.G.},
  \bibinfo{year}{2021}.
\newblock \bibinfo{title}{Optimizing surveillance and management of emerald ash
  borer in urban environments}.
\newblock \bibinfo{journal}{Natural Resource Modeling} \bibinfo{volume}{34},
  \bibinfo{pages}{e12267}.
%Type = Article
\bibitem[{Büyüktahtakın and Haight(2018)}]{buyuktahtakin_review_2018}
\bibinfo{author}{Büyüktahtakın, .E.}, \bibinfo{author}{Haight, R.G.},
  \bibinfo{year}{2018}.
\newblock \bibinfo{title}{A review of operations research models in invasive
  species management: state of the art, challenges, and future directions}.
\newblock \bibinfo{journal}{Annals of Operations Research}
  \bibinfo{volume}{271}, \bibinfo{pages}{357--403}.
\newblock \URLprefix \url{http://link.springer.com/10.1007/s10479-017-2670-5},
  \DOIprefix\doi{10.1007/s10479-017-2670-5}.
%Type = Article
\bibitem[{Cade and Noon(2003)}]{cade2003gentle}
\bibinfo{author}{Cade, B.S.}, \bibinfo{author}{Noon, B.R.},
  \bibinfo{year}{2003}.
\newblock \bibinfo{title}{A gentle introduction to quantile regression for
  ecologists}.
\newblock \bibinfo{journal}{Frontiers in Ecology and the Environment}
  \bibinfo{volume}{1}, \bibinfo{pages}{412--420}.
%Type = Article
\bibitem[{Chades et~al.(2011)Chades, Martin, Nicol, Burgman, Possingham and
  Buckley}]{chades_general_2011}
\bibinfo{author}{Chades, I.}, \bibinfo{author}{Martin, T.G.},
  \bibinfo{author}{Nicol, S.}, \bibinfo{author}{Burgman, M.A.},
  \bibinfo{author}{Possingham, H.P.}, \bibinfo{author}{Buckley, Y.M.},
  \bibinfo{year}{2011}.
\newblock \bibinfo{title}{General rules for managing and surveying networks of
  pests, diseases, and endangered species}.
\newblock \bibinfo{journal}{Proceedings of the National Academy of Sciences}
  \bibinfo{volume}{108}, \bibinfo{pages}{8323--8328}.
\newblock \URLprefix \url{http://www.pnas.org/cgi/doi/10.1073/pnas.1016846108},
  \DOIprefix\doi{10.1073/pnas.1016846108}.
%Type = Article
\bibitem[{Chakraborti et~al.(2016)Chakraborti, Madon and
  Kaur}]{chakraborti2016costs}
\bibinfo{author}{Chakraborti, R.K.}, \bibinfo{author}{Madon, S.},
  \bibinfo{author}{Kaur, J.}, \bibinfo{year}{2016}.
\newblock \bibinfo{title}{Costs for controlling dreissenid mussels affecting
  drinking water infrastructure: Case studies}.
\newblock \bibinfo{journal}{Journal-American Water Works Association}
  \bibinfo{volume}{108}, \bibinfo{pages}{E442--E453}.
%Type = Incollection
\bibitem[{Charles and Dukes(2007)}]{charles_impacts_2007}
\bibinfo{author}{Charles, H.}, \bibinfo{author}{Dukes, J.S.},
  \bibinfo{year}{2007}.
\newblock \bibinfo{title}{Impacts of {Invasive} {Species} on {Ecosystem}
  {Services}}, in: \bibinfo{editor}{Nentwig, W.} (Ed.),
  \bibinfo{booktitle}{Biological {Invasions}}. \bibinfo{publisher}{Springer},
  \bibinfo{address}{Berlin, Heidelberg}. Ecological {Studies}, pp.
  \bibinfo{pages}{217--237}.
\newblock \URLprefix \url{https://doi.org/10.1007/978-3-540-36920-2_13},
  \DOIprefix\doi{10.1007/978-3-540-36920-2_13}.
%Type = Article
\bibitem[{Csardi and Nepusz(2006)}]{igraph2006}
\bibinfo{author}{Csardi, G.}, \bibinfo{author}{Nepusz, T.},
  \bibinfo{year}{2006}.
\newblock \bibinfo{title}{The igraph software package for complex network
  research}.
\newblock \bibinfo{journal}{InterJournal} \bibinfo{volume}{Complex Systems},
  \bibinfo{pages}{1695}.
\newblock \URLprefix \url{https://igraph.org}.
%Type = Article
\bibitem[{Diagne et~al.(2021)Diagne, Leroy, Vaissière, Gozlan, Roiz, Jarić,
  Salles, Bradshaw and Courchamp}]{diagne_high_2021}
\bibinfo{author}{Diagne, C.}, \bibinfo{author}{Leroy, B.},
  \bibinfo{author}{Vaissière, A.C.}, \bibinfo{author}{Gozlan, R.E.},
  \bibinfo{author}{Roiz, D.}, \bibinfo{author}{Jarić, I.},
  \bibinfo{author}{Salles, J.M.}, \bibinfo{author}{Bradshaw, C.J.A.},
  \bibinfo{author}{Courchamp, F.}, \bibinfo{year}{2021}.
\newblock \bibinfo{title}{High and rising economic costs of biological
  invasions worldwide}.
\newblock \bibinfo{journal}{Nature} \bibinfo{volume}{592},
  \bibinfo{pages}{571--576}.
\newblock \URLprefix \url{https://www.nature.com/articles/s41586-021-03405-6},
  \DOIprefix\doi{10.1038/s41586-021-03405-6}.
%Type = Article
\bibitem[{Epanchin-Niell et~al.(2009)Epanchin-Niell, Hufford, Aslan, Sexton,
  Port and Waring}]{epanchin-niell_complex_landscapes_2009}
\bibinfo{author}{Epanchin-Niell, R.}, \bibinfo{author}{Hufford, M.},
  \bibinfo{author}{Aslan, C.}, \bibinfo{author}{Sexton, J.},
  \bibinfo{author}{Port, J.}, \bibinfo{author}{Waring, T.},
  \bibinfo{year}{2009}.
\newblock \bibinfo{title}{Controlling invasive species in complex social
  landscapes}.
\newblock \bibinfo{journal}{Frontiers in Ecology and the Environment}
  \bibinfo{volume}{8}, \bibinfo{pages}{210--216}.
\newblock \DOIprefix\doi{10.1890/090029}.
%Type = Article
\bibitem[{Epanchin-Niell(2017)}]{epanchin-niell_economics_2017}
\bibinfo{author}{Epanchin-Niell, R.S.}, \bibinfo{year}{2017}.
\newblock \bibinfo{title}{Economics of invasive species policy and management}.
\newblock \bibinfo{journal}{Biological Invasions} \bibinfo{volume}{19},
  \bibinfo{pages}{3333--3354}.
\newblock \URLprefix \url{http://link.springer.com/10.1007/s10530-017-1406-4},
  \DOIprefix\doi{10.1007/s10530-017-1406-4}.
%Type = Article
\bibitem[{Epanchin-Niell and Hastings(2010)}]{epanchin-niell_controlling_2010}
\bibinfo{author}{Epanchin-Niell, R.S.}, \bibinfo{author}{Hastings, A.},
  \bibinfo{year}{2010}.
\newblock \bibinfo{title}{Controlling established invaders: integrating
  economics and spread dynamics to determine optimal management}.
\newblock \bibinfo{journal}{Ecology Letters} \bibinfo{volume}{13},
  \bibinfo{pages}{528--541}.
\newblock \URLprefix
  \url{http://doi.wiley.com/10.1111/j.1461-0248.2010.01440.x},
  \DOIprefix\doi{10.1111/j.1461-0248.2010.01440.x}.
%Type = Article
\bibitem[{Epanchin-Niell and Wilen(2012)}]{epanchin-niell_optimal_2012}
\bibinfo{author}{Epanchin-Niell, R.S.}, \bibinfo{author}{Wilen, J.E.},
  \bibinfo{year}{2012}.
\newblock \bibinfo{title}{Optimal spatial control of biological invasions}.
\newblock \bibinfo{journal}{Journal of Environmental Economics and Management}
  \bibinfo{volume}{63}, \bibinfo{pages}{260--270}.
\newblock \URLprefix
  \url{https://www.sciencedirect.com/science/article/pii/S0095069611001392},
  \DOIprefix\doi{10.1016/j.jeem.2011.10.003}.
%Type = Article
\bibitem[{Epanchin-Niell and Wilen(2015)}]{epanchin-niell_individual_2015}
\bibinfo{author}{Epanchin-Niell, R.S.}, \bibinfo{author}{Wilen, J.E.},
  \bibinfo{year}{2015}.
\newblock \bibinfo{title}{Individual and {Cooperative} {Management} of
  {Invasive} {Species} in {Human}-mediated {Landscapes}}.
\newblock \bibinfo{journal}{American Journal of Agricultural Economics}
  \bibinfo{volume}{97}, \bibinfo{pages}{180--198}.
\newblock \URLprefix
  \url{https://ideas.repec.org/a/oup/ajagec/v97y2015i1p180-198..html}.
%Type = Article
\bibitem[{Erkol et~al.(2019)Erkol, Castellano and
  Radicchi}]{erkol2019systematic}
\bibinfo{author}{Erkol, {\c{S}}.}, \bibinfo{author}{Castellano, C.},
  \bibinfo{author}{Radicchi, F.}, \bibinfo{year}{2019}.
\newblock \bibinfo{title}{Systematic comparison between methods for the
  detection of influential spreaders in complex networks}.
\newblock \bibinfo{journal}{Scientific reports} \bibinfo{volume}{9},
  \bibinfo{pages}{1--11}.
%Type = Article
\bibitem[{Feige(1998)}]{feige1998threshold}
\bibinfo{author}{Feige, U.}, \bibinfo{year}{1998}.
\newblock \bibinfo{title}{A threshold of ln n for approximating set cover}.
\newblock \bibinfo{journal}{Journal of the ACM (JACM)} \bibinfo{volume}{45},
  \bibinfo{pages}{634--652}.
%Type = Article
\bibitem[{Fischer et~al.(2021)Fischer, Beck, Herborg and
  Lewis}]{fischer2021managing}
\bibinfo{author}{Fischer, S.M.}, \bibinfo{author}{Beck, M.},
  \bibinfo{author}{Herborg, L.M.}, \bibinfo{author}{Lewis, M.A.},
  \bibinfo{year}{2021}.
\newblock \bibinfo{title}{Managing aquatic invasions: optimal locations and
  operating times for watercraft inspection stations}.
\newblock \bibinfo{journal}{Journal of Environmental Management}
  \bibinfo{volume}{283}, \bibinfo{pages}{111923}.
%Type = Article
\bibitem[{Freeman(1978)}]{freeman1978centrality}
\bibinfo{author}{Freeman, L.C.}, \bibinfo{year}{1978}.
\newblock \bibinfo{title}{Centrality in social networks conceptual
  clarification}.
\newblock \bibinfo{journal}{Social networks} \bibinfo{volume}{1},
  \bibinfo{pages}{215--239}.
%Type = Article
\bibitem[{de~la Fuente et~al.(2018)de~la Fuente, Saura and
  Beck}]{de2018predicting}
\bibinfo{author}{de~la Fuente, B.}, \bibinfo{author}{Saura, S.},
  \bibinfo{author}{Beck, P.S.}, \bibinfo{year}{2018}.
\newblock \bibinfo{title}{Predicting the spread of an invasive tree pest: the
  pine wood nematode in southern europe}.
\newblock \bibinfo{journal}{Journal of Applied Ecology} \bibinfo{volume}{55},
  \bibinfo{pages}{2374--2385}.
%Type = Article
\bibitem[{Gallardo et~al.(2016)Gallardo, Clavero, Sánchez and
  Vilà}]{gallardo_global_2016}
\bibinfo{author}{Gallardo, B.}, \bibinfo{author}{Clavero, M.},
  \bibinfo{author}{Sánchez, M.}, \bibinfo{author}{Vilà, M.},
  \bibinfo{year}{2016}.
\newblock \bibinfo{title}{Global ecological impacts of invasive species in
  aquatic ecosystems.}
\newblock \bibinfo{journal}{Global Change Biology} \bibinfo{volume}{22},
  \bibinfo{pages}{151--163}.
\newblock \DOIprefix\doi{10.1111/gcb.13004}.
%Type = Article
\bibitem[{Haight et~al.(2021)Haight, Kinsley, Kao, Yemshanov and
  Phelps}]{haight2021paper}
\bibinfo{author}{Haight, R.G.}, \bibinfo{author}{Kinsley, A.C.},
  \bibinfo{author}{Kao, S.Y.}, \bibinfo{author}{Yemshanov, D.},
  \bibinfo{author}{Phelps, N.B.}, \bibinfo{year}{2021}.
\newblock \bibinfo{title}{Optimizing the location of watercraft inspection
  stations to slow the spread of aquatic invasive species}.
\newblock \bibinfo{journal}{Biological Invasions} \bibinfo{volume}{23},
  \bibinfo{pages}{3907–--3919}.
%Type = Article
\bibitem[{Holme et~al.(2002)Holme, Kim, Yoon and Han}]{holme_attack_2002}
\bibinfo{author}{Holme, P.}, \bibinfo{author}{Kim, B.J.},
  \bibinfo{author}{Yoon, C.N.}, \bibinfo{author}{Han, S.K.},
  \bibinfo{year}{2002}.
\newblock \bibinfo{title}{Attack vulnerability of complex networks}.
\newblock \bibinfo{journal}{Physical Review E} \bibinfo{volume}{65},
  \bibinfo{pages}{056109}.
\newblock \URLprefix \url{https://link.aps.org/doi/10.1103/PhysRevE.65.056109},
  \DOIprefix\doi{10.1103/PhysRevE.65.056109}.
%Type = Article
\bibitem[{Johnson et~al.(2001)Johnson, Ricciardi and
  Carlton}]{johnson_overland_2001}
\bibinfo{author}{Johnson, L.E.}, \bibinfo{author}{Ricciardi, A.},
  \bibinfo{author}{Carlton, J.T.}, \bibinfo{year}{2001}.
\newblock \bibinfo{title}{Overland {Dispersal} of {Aquatic} {Invasive}
  {Species}: {A} {Risk} {Assessment} of {Transient} {Recreational} {Boating}}.
\newblock \bibinfo{journal}{Ecological Applications} \bibinfo{volume}{11},
  \bibinfo{pages}{1789--1799}.
%Type = Article
\bibitem[{Kanankege et~al.(2018)Kanankege, Alkhamis, Phelps and
  Perez}]{kanankege2018probability}
\bibinfo{author}{Kanankege, K.S.}, \bibinfo{author}{Alkhamis, M.A.},
  \bibinfo{author}{Phelps, N.B.}, \bibinfo{author}{Perez, A.M.},
  \bibinfo{year}{2018}.
\newblock \bibinfo{title}{A probability co-kriging model to account for
  reporting bias and recognize areas at high risk for zebra mussels and
  eurasian watermilfoil invasions in minnesota}.
\newblock \bibinfo{journal}{Frontiers in Veterinary Science}
  \bibinfo{volume}{4}, \bibinfo{pages}{231}.
%Type = Article
\bibitem[{Kanankege et~al.(2020)Kanankege, Phelps, Vesterinen, Errecaborde,
  Alvarez, Bender, Wells and Perez}]{kanankege2020lessons}
\bibinfo{author}{Kanankege, K.S.}, \bibinfo{author}{Phelps, N.B.},
  \bibinfo{author}{Vesterinen, H.M.}, \bibinfo{author}{Errecaborde, K.M.},
  \bibinfo{author}{Alvarez, J.}, \bibinfo{author}{Bender, J.B.},
  \bibinfo{author}{Wells, S.J.}, \bibinfo{author}{Perez, A.M.},
  \bibinfo{year}{2020}.
\newblock \bibinfo{title}{Lessons learned from the stakeholder engagement in
  research: application of spatial analytical tools in one health problems}.
\newblock \bibinfo{journal}{Frontiers in Veterinary Science}
  \bibinfo{volume}{7}, \bibinfo{pages}{254}.
%Type = Misc
\bibitem[{Kao et~al.(2020)Kao, Enns, Tomamichel, Doll, Escobar, Qiao, Craft and
  Phelps}]{kao2020}
\bibinfo{author}{Kao, S.Y.}, \bibinfo{author}{Enns, E.A.},
  \bibinfo{author}{Tomamichel, M.}, \bibinfo{author}{Doll, A.},
  \bibinfo{author}{Escobar, L.E.}, \bibinfo{author}{Qiao, H.},
  \bibinfo{author}{Craft, M.E.}, \bibinfo{author}{Phelps, N.B.D.},
  \bibinfo{year}{2020}.
\newblock \bibinfo{title}{Network connectivity patterns of {Minnesota}
  waterbodies and implications for aquatic invasive species prevention}.
\newblock \bibinfo{howpublished}{Retrieved from the Data Repository for the
  University of Minnesota: https://doi.org/10.13020/DJW8-2V86}.
%Type = Article
\bibitem[{Kao et~al.(2021)Kao, Enns, Tomamichel, Doll, Escobar, Qiao, Craft and
  Phelps}]{kao2021paper}
\bibinfo{author}{Kao, S.Y.Z.}, \bibinfo{author}{Enns, E.A.},
  \bibinfo{author}{Tomamichel, M.}, \bibinfo{author}{Doll, A.},
  \bibinfo{author}{Escobar, L.E.}, \bibinfo{author}{Qiao, H.},
  \bibinfo{author}{Craft, M.E.}, \bibinfo{author}{Phelps, N.B.},
  \bibinfo{year}{2021}.
\newblock \bibinfo{title}{Network connectivity of {Minnesota} waterbodies and
  implications for aquatic invasive species prevention}.
\newblock \bibinfo{journal}{Biological Invasions} \bibinfo{volume}{23},
  \bibinfo{pages}{3231–--3242}.
%Type = Incollection
\bibitem[{Karatayev et~al.(2002)Karatayev, Burlakova and
  Padilla}]{karatayev2002impacts}
\bibinfo{author}{Karatayev, A.Y.}, \bibinfo{author}{Burlakova, L.E.},
  \bibinfo{author}{Padilla, D.K.}, \bibinfo{year}{2002}.
\newblock \bibinfo{title}{Impacts of zebra mussels on aquatic communities and
  their role as ecosystem engineers}, in: \bibinfo{booktitle}{Invasive aquatic
  species of Europe. Distribution, impacts and management}.
  \bibinfo{publisher}{Springer}, pp. \bibinfo{pages}{433--446}.
%Type = Inproceedings
\bibitem[{Kempe et~al.(2003)Kempe, Kleinberg and
  Tardos}]{kempe_maximizing_2003}
\bibinfo{author}{Kempe, D.}, \bibinfo{author}{Kleinberg, J.},
  \bibinfo{author}{Tardos, E.}, \bibinfo{year}{2003}.
\newblock \bibinfo{title}{Maximizing the spread of influence through a social
  network}, in: \bibinfo{booktitle}{Proceedings of the ninth {ACM} {SIGKDD}
  international conference on {Knowledge} discovery and data mining - {KDD}
  '03}, \bibinfo{publisher}{ACM Press}, \bibinfo{address}{Washington, D.C.}. p.
  \bibinfo{pages}{137}.
\newblock \URLprefix
  \url{http://portal.acm.org/citation.cfm?doid=956750.956769},
  \DOIprefix\doi{10.1145/956750.956769}.
%Type = Article
\bibitem[{Kinsley et~al.(2021)Kinsley, Haight, Snellgrove, Muellner, Muellner,
  Duhr and Phelps}]{kinseyinrev}
\bibinfo{author}{Kinsley, A.C.}, \bibinfo{author}{Haight, R.G.},
  \bibinfo{author}{Snellgrove, N.}, \bibinfo{author}{Muellner, P.},
  \bibinfo{author}{Muellner, U.}, \bibinfo{author}{Duhr, M.},
  \bibinfo{author}{Phelps, N.B.D.}, \bibinfo{year}{2021}.
\newblock \bibinfo{title}{{AIS Explorer}: {Prioritization} for watercraft
  inspections-{A decision-support tool for aquatic invasive species
  management}}.
\newblock \bibinfo{journal}{arXiv:2112.03859 [math, q-bio]} \URLprefix
  \url{https://arxiv.org/abs/2112.03859}. \bibinfo{note}{arXiv: 2112.03859}.
%Type = Inproceedings
\bibitem[{Kleinberg(1998)}]{kleinberg1998authoritative}
\bibinfo{author}{Kleinberg, J.M.}, \bibinfo{year}{1998}.
\newblock \bibinfo{title}{Authoritative sources in a hyperlinked environment},
  in: \bibinfo{booktitle}{Proceedings of the ninth annual ACM-SIAM symposium on
  Discrete algorithms}, pp. \bibinfo{pages}{668--677}.
%Type = Article
\bibitem[{Kleinberg(1999)}]{kleinberg1999authoritative}
\bibinfo{author}{Kleinberg, J.M.}, \bibinfo{year}{1999}.
\newblock \bibinfo{title}{Authoritative sources in a hyperlinked environment}.
\newblock \bibinfo{journal}{Journal of the {ACM}} \bibinfo{volume}{46},
  \bibinfo{pages}{604--632}.
%Type = Incollection
\bibitem[{Koenker(1994)}]{koenker1994confidence}
\bibinfo{author}{Koenker, R.}, \bibinfo{year}{1994}.
\newblock \bibinfo{title}{Confidence intervals for regression quantiles}, in:
  \bibinfo{booktitle}{Asymptotic Statistics}. \bibinfo{publisher}{Springer},
  pp. \bibinfo{pages}{349--359}.
%Type = Article
\bibitem[{Kroetz and Sanchirico(2015)}]{kroetz2015bioeconomics}
\bibinfo{author}{Kroetz, K.}, \bibinfo{author}{Sanchirico, J.},
  \bibinfo{year}{2015}.
\newblock \bibinfo{title}{The bioeconomics of spatial-dynamic systems in
  natural resource management}.
\newblock \bibinfo{journal}{Annual Review of Resource Economics}
  \bibinfo{volume}{7}, \bibinfo{pages}{189--207}.
%Type = Article
\bibitem[{Kvistad et~al.(2019)Kvistad, Chadderton and
  Bossenbroek}]{kvistad2019network}
\bibinfo{author}{Kvistad, J.T.}, \bibinfo{author}{Chadderton, W.L.},
  \bibinfo{author}{Bossenbroek, J.M.}, \bibinfo{year}{2019}.
\newblock \bibinfo{title}{Network centrality as a potential method for
  prioritizing ports for aquatic invasive species surveillance and response in
  the {Laurentian Great Lakes}}.
\newblock \bibinfo{journal}{Management of Biological Invasions}
  \bibinfo{volume}{10}, \bibinfo{pages}{403}.
%Type = Article
\bibitem[{Leung et~al.(2006)Leung, Bossenbroek and Lodge}]{leung_boats_2006}
\bibinfo{author}{Leung, B.}, \bibinfo{author}{Bossenbroek, J.M.},
  \bibinfo{author}{Lodge, D.M.}, \bibinfo{year}{2006}.
\newblock \bibinfo{title}{Boats, {Pathways}, and {Aquatic} {Biological}
  {Invasions}: {Estimating} {Dispersal} {Potential} with {Gravity} {Models}}.
\newblock \bibinfo{journal}{Biological Invasions} \bibinfo{volume}{8},
  \bibinfo{pages}{241--254}.
\newblock \URLprefix \url{http://link.springer.com/10.1007/s10530-004-5573-8},
  \DOIprefix\doi{10.1007/s10530-004-5573-8}.
%Type = Article
\bibitem[{Leung et~al.(2002)Leung, Lodge, Finnoff, Shogren, Lewis and
  Lamberti}]{leung_ounce_2002}
\bibinfo{author}{Leung, B.}, \bibinfo{author}{Lodge, D.M.},
  \bibinfo{author}{Finnoff, D.}, \bibinfo{author}{Shogren, J.F.},
  \bibinfo{author}{Lewis, M.A.}, \bibinfo{author}{Lamberti, G.},
  \bibinfo{year}{2002}.
\newblock \bibinfo{title}{An ounce of prevention or a pound of cure:
  bioeconomic risk analysis of invasive species}.
\newblock \bibinfo{journal}{Proceedings of the Royal Society of London. Series
  B: Biological Sciences} \bibinfo{volume}{269}, \bibinfo{pages}{2407--2413}.
\newblock \URLprefix
  \url{https://royalsocietypublishing.org/doi/10.1098/rspb.2002.2179},
  \DOIprefix\doi{10.1098/rspb.2002.2179}.
%Type = Article
\bibitem[{Lund et~al.(2018)Lund, Cattoor, Fieldseth, Sweet and
  McCartney}]{lund2018zebra}
\bibinfo{author}{Lund, K.}, \bibinfo{author}{Cattoor, K.B.},
  \bibinfo{author}{Fieldseth, E.}, \bibinfo{author}{Sweet, J.},
  \bibinfo{author}{McCartney, M.A.}, \bibinfo{year}{2018}.
\newblock \bibinfo{title}{Zebra mussel (\textit{Dreissena polymorpha})
  eradication efforts in {Christmas Lake, Minnesota}}.
\newblock \bibinfo{journal}{Lake and Reservoir Management}
  \bibinfo{volume}{34}, \bibinfo{pages}{7--20}.
%Type = Article
\bibitem[{Mallez and McCartney(2018)}]{mallez2018dispersal}
\bibinfo{author}{Mallez, S.}, \bibinfo{author}{McCartney, M.},
  \bibinfo{year}{2018}.
\newblock \bibinfo{title}{Dispersal mechanisms for zebra mussels: population
  genetics supports clustered invasions over spread from hub lakes in
  minnesota}.
\newblock \bibinfo{journal}{Biological Invasions} \bibinfo{volume}{20},
  \bibinfo{pages}{2461--2484}.
%Type = Article
\bibitem[{McDonald-Madden et~al.(2016)McDonald-Madden, Sabbadin, Game, Baxter,
  Chadès and Possingham}]{mcdonald-madden_using_2016}
\bibinfo{author}{McDonald-Madden, E.}, \bibinfo{author}{Sabbadin, R.},
  \bibinfo{author}{Game, E.T.}, \bibinfo{author}{Baxter, P.W.J.},
  \bibinfo{author}{Chadès, I.}, \bibinfo{author}{Possingham, H.P.},
  \bibinfo{year}{2016}.
\newblock \bibinfo{title}{Using food-web theory to conserve ecosystems}.
\newblock \bibinfo{journal}{Nature Communications} \bibinfo{volume}{7},
  \bibinfo{pages}{10245}.
\newblock \URLprefix \url{http://www.nature.com/articles/ncomms10245},
  \DOIprefix\doi{10.1038/ncomms10245}.
%Type = Article
\bibitem[{McEachran et~al.(2019)McEachran, Trapp, Zimmer, Herwig, Hegedus,
  Herzog and Staples}]{mceachran2019stable}
\bibinfo{author}{McEachran, M.C.}, \bibinfo{author}{Trapp, R.S.},
  \bibinfo{author}{Zimmer, K.D.}, \bibinfo{author}{Herwig, B.R.},
  \bibinfo{author}{Hegedus, C.E.}, \bibinfo{author}{Herzog, C.E.},
  \bibinfo{author}{Staples, D.F.}, \bibinfo{year}{2019}.
\newblock \bibinfo{title}{Stable isotopes indicate that zebra mussels
  (dreissena polymorpha) increase dependence of lake food webs on littoral
  energy sources}.
\newblock \bibinfo{journal}{Freshwater Biology} \bibinfo{volume}{64},
  \bibinfo{pages}{183--196}.
%Type = Book
\bibitem[{{Minnesota Department of Natural Resources}(2020)}]{MNDNR2020}
\bibinfo{author}{{Minnesota Department of Natural Resources}},
  \bibinfo{year}{2020}.
\newblock \bibinfo{title}{{Invasive Species of Aquatic Plants} and {Wild
  Animals} in {Minnesota}: {Annual Report for 2020}}.
\newblock \bibinfo{publisher}{State of Minnesota, Department of Natural
  Resources}, \bibinfo{address}{St. Paul, MN}.
%Type = Article
\bibitem[{Minor and Urban(2008)}]{minor_graph-theory_2008}
\bibinfo{author}{Minor, E.S.}, \bibinfo{author}{Urban, D.L.},
  \bibinfo{year}{2008}.
\newblock \bibinfo{title}{A {Graph}-{Theory} {Framework} for {Evaluating}
  {Landscape} {Connectivity} and {Conservation} {Planning}: \textit{{Graph}
  {Theory}, {Connectivity}, and {Conservation}}}.
\newblock \bibinfo{journal}{Conservation Biology} \bibinfo{volume}{22},
  \bibinfo{pages}{297--307}.
\newblock \URLprefix
  \url{http://doi.wiley.com/10.1111/j.1523-1739.2007.00871.x},
  \DOIprefix\doi{10.1111/j.1523-1739.2007.00871.x}.
%Type = Article
\bibitem[{Morel‐Journel et~al.(2019)Morel‐Journel, Assa, Mailleret and
  Vercken}]{moreljournel_its_2018}
\bibinfo{author}{Morel‐Journel, T.}, \bibinfo{author}{Assa, C.R.},
  \bibinfo{author}{Mailleret, L.}, \bibinfo{author}{Vercken, E.},
  \bibinfo{year}{2019}.
\newblock \bibinfo{title}{Its all about connections: hubs and invasion in
  habitat networks}.
\newblock \bibinfo{journal}{Ecology Letters} \bibinfo{volume}{22},
  \bibinfo{pages}{313--321}.
\newblock \URLprefix
  \url{https://onlinelibrary.wiley.com/doi/abs/10.1111/ele.13192},
  \DOIprefix\doi{10.1111/ele.13192}.
%Type = Article
\bibitem[{Muirhead and Macisaac(2005)}]{muirhead_development_2005}
\bibinfo{author}{Muirhead, J.R.}, \bibinfo{author}{Macisaac, H.J.},
  \bibinfo{year}{2005}.
\newblock \bibinfo{title}{Development of inland lakes as hubs in an invasion
  network}.
\newblock \bibinfo{journal}{Journal of Applied Ecology} \bibinfo{volume}{42},
  \bibinfo{pages}{80--90}.
\newblock \URLprefix
  \url{http://doi.wiley.com/10.1111/j.1365-2664.2004.00988.x},
  \DOIprefix\doi{10.1111/j.1365-2664.2004.00988.x}.
%Type = Article
\bibitem[{Newman(2002)}]{newman_spread_2002}
\bibinfo{author}{Newman, M.E.J.}, \bibinfo{year}{2002}.
\newblock \bibinfo{title}{Spread of epidemic disease on networks}.
\newblock \bibinfo{journal}{Physical Review E} \bibinfo{volume}{66},
  \bibinfo{pages}{016128}.
\newblock \URLprefix \url{https://link.aps.org/doi/10.1103/PhysRevE.66.016128},
  \DOIprefix\doi{10.1103/PhysRevE.66.016128}.
%Type = Article
\bibitem[{Nowzari et~al.(2016)Nowzari, Precaido and
  Pappas}]{nowzari_analysis_2016}
\bibinfo{author}{Nowzari, C.}, \bibinfo{author}{Precaido, V.M.},
  \bibinfo{author}{Pappas, G.J.}, \bibinfo{year}{2016}.
\newblock \bibinfo{title}{Analysis and {Control} of {Epidemics}: {A} {Survey}
  of {Spreading} {Processes} on {Complex} {Networks}}.
\newblock \bibinfo{journal}{IEEE Control Systems} \bibinfo{volume}{36},
  \bibinfo{pages}{26--46}.
\newblock \URLprefix \url{https://ieeexplore.ieee.org/document/7393962/},
  \DOIprefix\doi{10.1109/MCS.2015.2495000}.
%Type = Article
\bibitem[{Pastor-Satorras et~al.(2015)Pastor-Satorras, Castellano, Van~Mieghem
  and Vespignani}]{pastor-satorras_epidemic_2015}
\bibinfo{author}{Pastor-Satorras, R.}, \bibinfo{author}{Castellano, C.},
  \bibinfo{author}{Van~Mieghem, P.}, \bibinfo{author}{Vespignani, A.},
  \bibinfo{year}{2015}.
\newblock \bibinfo{title}{Epidemic processes in complex networks}.
\newblock \bibinfo{journal}{Reviews of Modern Physics} \bibinfo{volume}{87},
  \bibinfo{pages}{925--979}.
\newblock \URLprefix \url{https://link.aps.org/doi/10.1103/RevModPhys.87.925},
  \DOIprefix\doi{10.1103/RevModPhys.87.925}.
%Type = Article
\bibitem[{Pastor-Satorras and
  Vespignani(2002)}]{pastor-satorras_immunization_2002}
\bibinfo{author}{Pastor-Satorras, R.}, \bibinfo{author}{Vespignani, A.},
  \bibinfo{year}{2002}.
\newblock \bibinfo{title}{Immunization of complex networks}.
\newblock \bibinfo{journal}{Physical Review E} \bibinfo{volume}{65},
  \bibinfo{pages}{036104}.
\newblock \URLprefix \url{https://link.aps.org/doi/10.1103/PhysRevE.65.036104},
  \DOIprefix\doi{10.1103/PhysRevE.65.036104}.
%Type = Article
\bibitem[{Pei et~al.(2018)Pei, Morone and Makse}]{pei_theories_2018}
\bibinfo{author}{Pei, S.}, \bibinfo{author}{Morone, F.},
  \bibinfo{author}{Makse, H.A.}, \bibinfo{year}{2018}.
\newblock \bibinfo{title}{Theories for influencer identification in complex
  networks}.
\newblock \bibinfo{journal}{arXiv:1707.01594 [physics]} \URLprefix
  \url{http://arxiv.org/abs/1707.01594}. \bibinfo{note}{arXiv: 1707.01594}.
%Type = Article
\bibitem[{Perry et~al.(2017)Perry, Moloney and Etherington}]{perry_using_2017}
\bibinfo{author}{Perry, G.L.W.}, \bibinfo{author}{Moloney, K.A.},
  \bibinfo{author}{Etherington, T.R.}, \bibinfo{year}{2017}.
\newblock \bibinfo{title}{Using network connectivity to prioritise sites for
  the control of invasive species}.
\newblock \bibinfo{journal}{Journal of Applied Ecology} \bibinfo{volume}{54},
  \bibinfo{pages}{1238--1250}.
\newblock \URLprefix \url{http://doi.wiley.com/10.1111/1365-2664.12827},
  \DOIprefix\doi{10.1111/1365-2664.12827}.
%Type = Inproceedings
\bibitem[{Prescott et~al.(2013)Prescott, Claudi and
  Prescott}]{prescott2013impact}
\bibinfo{author}{Prescott, T.H.}, \bibinfo{author}{Claudi, R.},
  \bibinfo{author}{Prescott, K.L.}, \bibinfo{year}{2013}.
\newblock \bibinfo{title}{Impact of {Dreissenid} mussels on the infrastructure
  of dams and hydroelectric power plants}, in: \bibinfo{booktitle}{{Quagga and
  Zebra Mussels}}, \bibinfo{publisher}{CRC Press}. pp.
  \bibinfo{pages}{243--258}.
%Type = Article
\bibitem[{Runting et~al.(2019)Runting, Griscom, Struebig, Satar, Meijaard,
  Burivalova, Cheyne, Deere, Game, Putz et~al.}]{runting2019larger}
\bibinfo{author}{Runting, R.K.}, \bibinfo{author}{Griscom, B.W.},
  \bibinfo{author}{Struebig, M.J.}, \bibinfo{author}{Satar, M.},
  \bibinfo{author}{Meijaard, E.}, \bibinfo{author}{Burivalova, Z.},
  \bibinfo{author}{Cheyne, S.M.}, \bibinfo{author}{Deere, N.J.},
  \bibinfo{author}{Game, E.T.}, \bibinfo{author}{Putz, F.E.}, et~al.,
  \bibinfo{year}{2019}.
\newblock \bibinfo{title}{Larger gains from improved management over
  sparing--sharing for tropical forests}.
\newblock \bibinfo{journal}{Nature Sustainability} \bibinfo{volume}{2},
  \bibinfo{pages}{53--61}.
%Type = Article
\bibitem[{Sardain et~al.(2019)Sardain, Sardain and Leung}]{sardain2019global}
\bibinfo{author}{Sardain, A.}, \bibinfo{author}{Sardain, E.},
  \bibinfo{author}{Leung, B.}, \bibinfo{year}{2019}.
\newblock \bibinfo{title}{Global forecasts of shipping traffic and biological
  invasions to 2050}.
\newblock \bibinfo{journal}{Nature Sustainability} \bibinfo{volume}{2},
  \bibinfo{pages}{274--282}.
%Type = Article
\bibitem[{Vander~Zanden and Olden(2008)}]{vander2008management}
\bibinfo{author}{Vander~Zanden, M.J.}, \bibinfo{author}{Olden, J.D.},
  \bibinfo{year}{2008}.
\newblock \bibinfo{title}{A management framework for preventing the secondary
  spread of aquatic invasive species}.
\newblock \bibinfo{journal}{Canadian Journal of Fisheries and Aquatic Sciences}
  \bibinfo{volume}{65}, \bibinfo{pages}{1512--1522}.

\end{thebibliography}

%%%%%%%%%%%%%%%%%%%%%%%%%%%%
\clearpage
\section*{Figures and Tables}
%%%%%%%%%%%%%%%%%%%%
\subsection*{Figures}

\newpage
\begin{figure}[H]
\centering
    \includegraphics{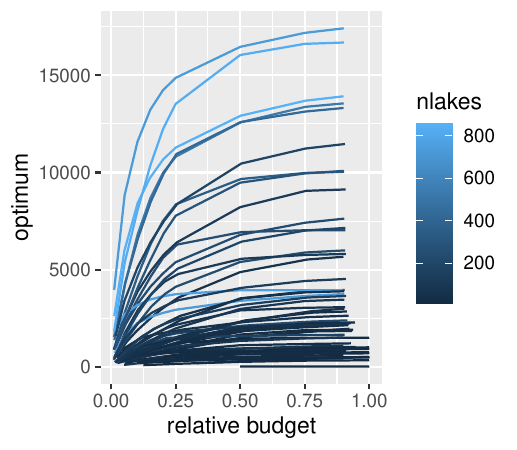}
\caption{\textbf{Optimum IP solution versus budget.} The y-axis shows the number of potentially infecting boats inspected in the IP solution with varying relative budget, for each county. Each line is a county, and lines are colored by the number of lakes in the county. Relative budget is the proportion of the budget required to inspect all infective boats in the county.}
\label{si:fig:dimret}
\end{figure}

\begin{figure*}[p]
\includegraphics[width=17.8cm]{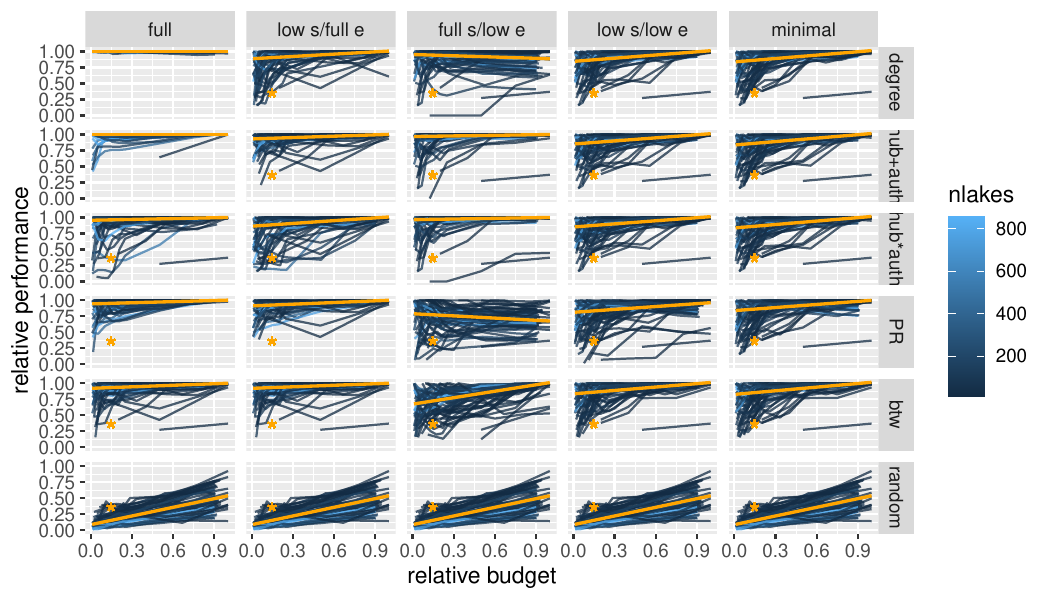}
\caption{\textbf{Relative performance of each metric-based
heuristic for the full and partial information scenarios (columns)
against relative budget for all metrics (rows).} Lines represent counties, colored by the number of lakes in the county. Orange lines denote the median (0.5 quantile) fitted by quantile regression. {\bf *} indicates significant regression slope of the 0.5 quantile (no overlap of 95\% confidence interval with 0).}\label{si:fig:budget}
\end{figure*}

\begin{figure*}[htp]
\includegraphics[width=15cm]{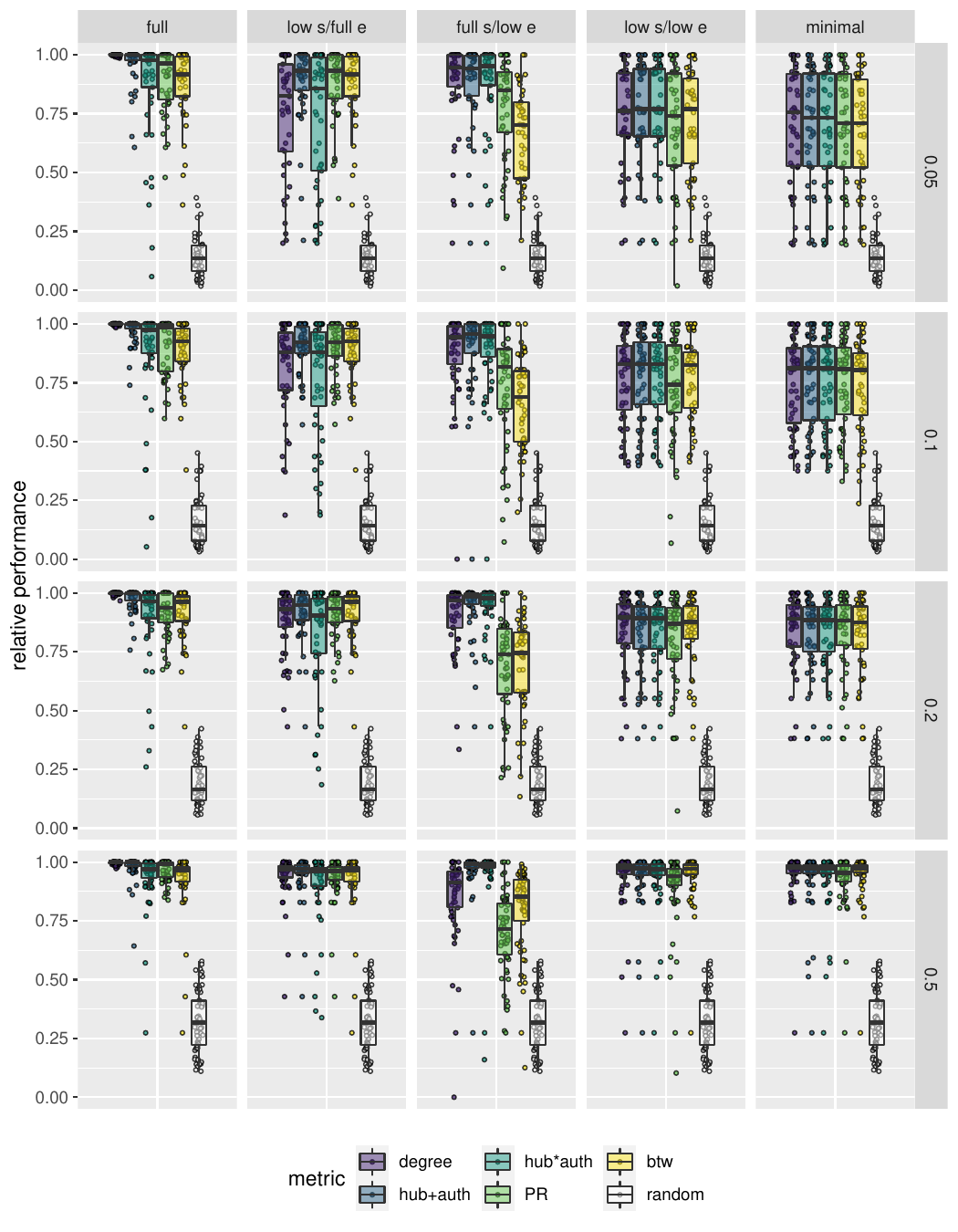}
\caption{\textbf{Boxplots of relative performance of each metric-based heuristic for the information levels (columns) and relative budget categories (rows).}
The distribution of performance across counties ($n=58$) is illustrated with boxplots (center line, median; box limits, upper and lower quartiles; whiskers, 1.5x interquartile range) overplotted on county values (circular points). Random inspection also is included for comparison; there, triangular points represent mean performance across many random inspection placements (See Methods). 
}
\label{si:fig:budget2}
\end{figure*}

\begin{figure*}
\includegraphics[width=16cm]{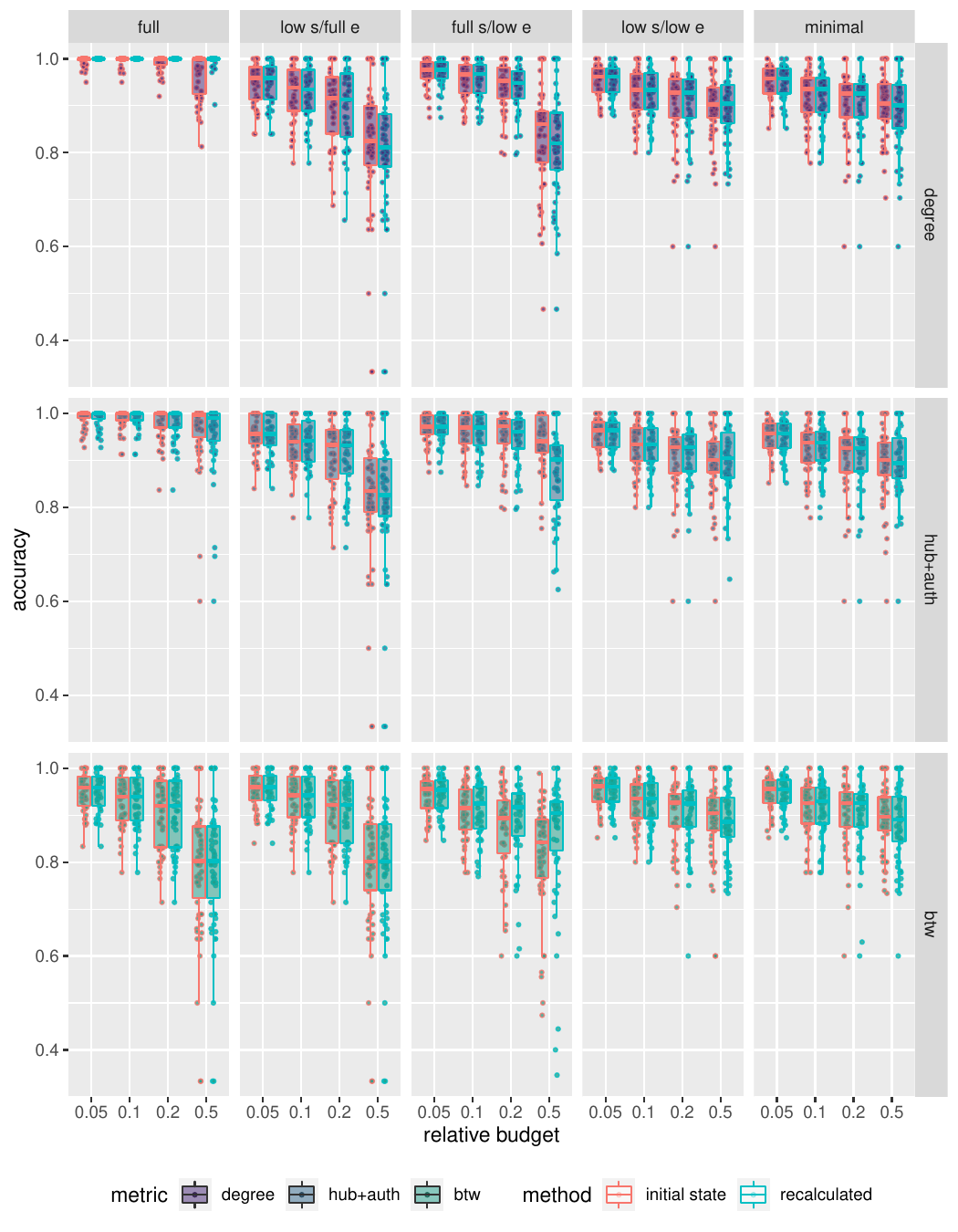}
\caption{\textbf{Boxplots of accuracy of each metric-based heuristic at matching the optimal inspection pattern for all information scenarios (columns) and several relative  budgets.}
Accuracy is the proportion of matching sites (inspected or uninspected) between the metric and optimal solution for a given budget.}
\label{si:fig:acc}
\end{figure*}

\begin{figure*}
\includegraphics{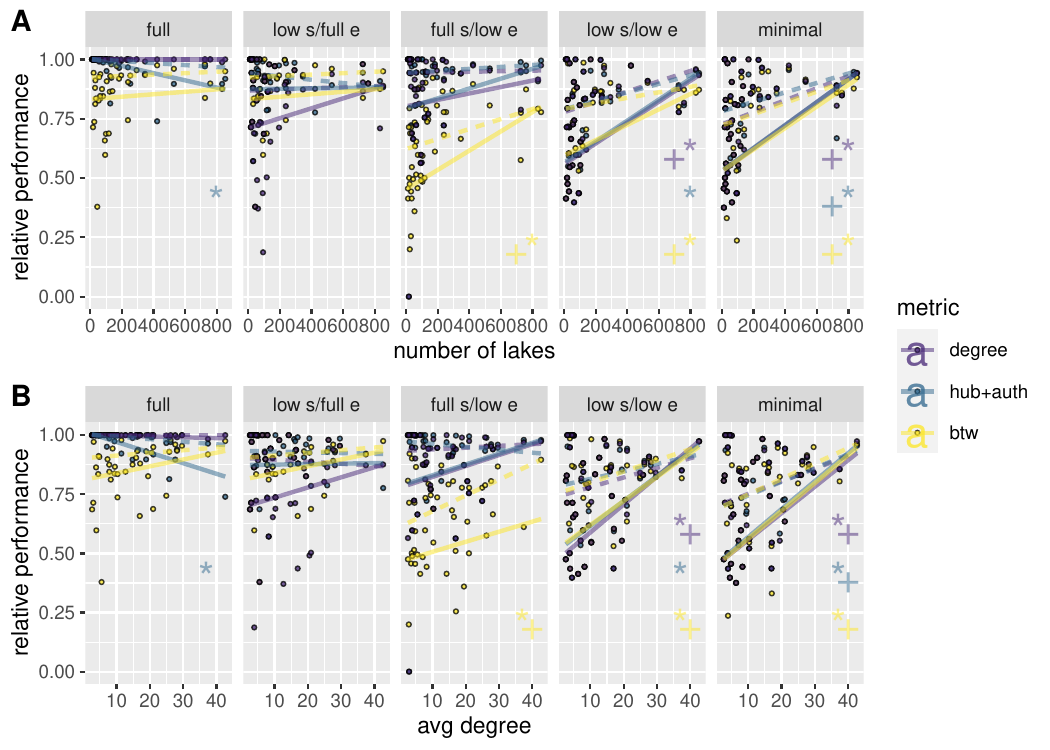}
\caption{\textbf{Relationship between performance and (A) network size or (B) network density (avg degree) across 58 counties (circular points).} Lines shown are quantile regression results for the 0.50 and 0.25 quantiles. Significance for the 0.25 (solid line) and 0.5 (dashed line) quantiles are marked by \textbf{*} and \textbf{+} respectively.}\label{si:fig:size-avgdeg}
\end{figure*}

\begin{figure*}
\centering
\includegraphics[width=16cm]{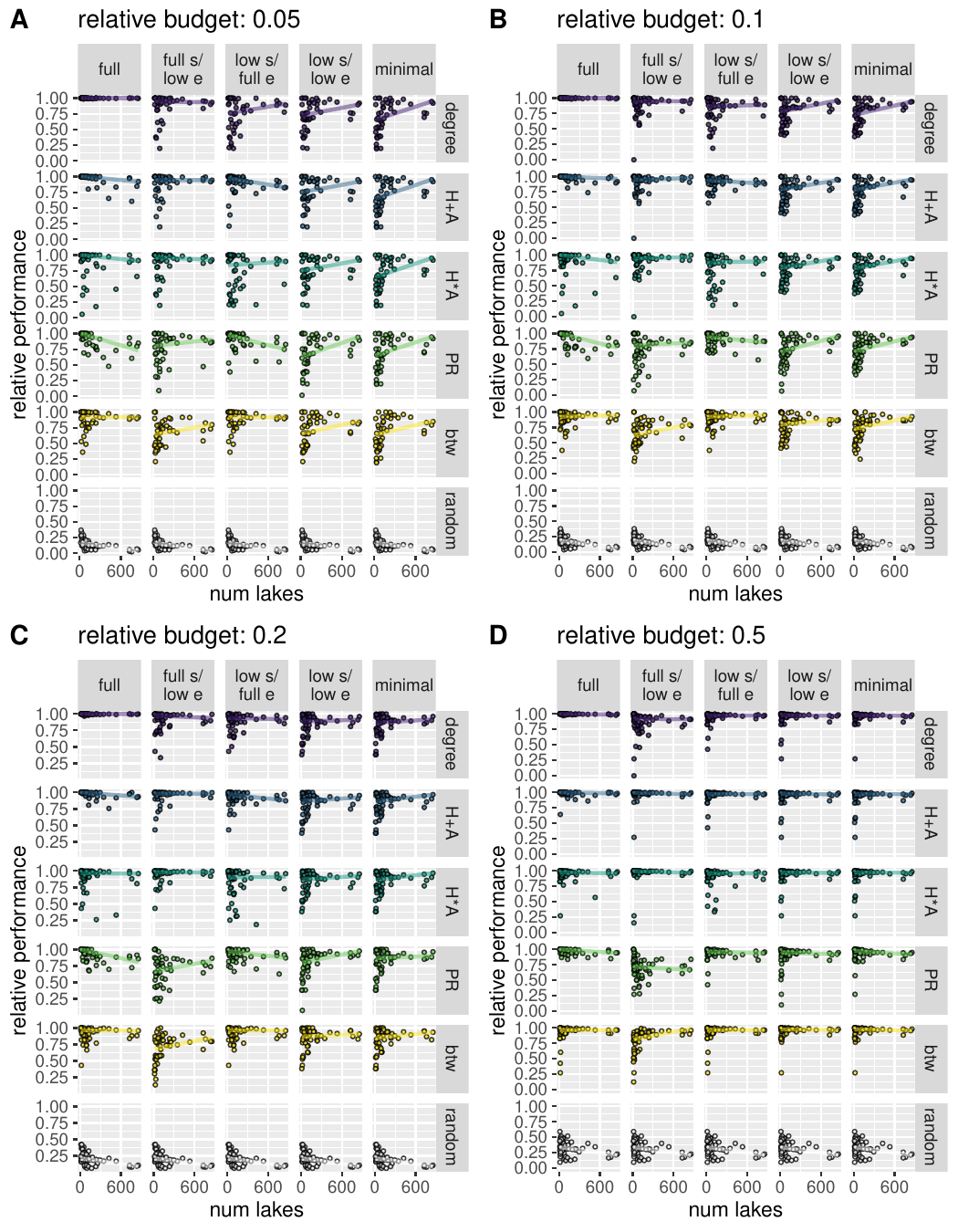}
\caption{\textbf{Relative performance versus network size for all information levels and all metrics for four example budgets.} We measure budget as a proportion of the total budget needed to inspect all infected boats. The four budgets explored here are: 0.05, 0.1, 0.2, and 0.5.}
\label{si:fig:size}
\end{figure*}

\begin{figure*}
\centering
\includegraphics[width=16cm]{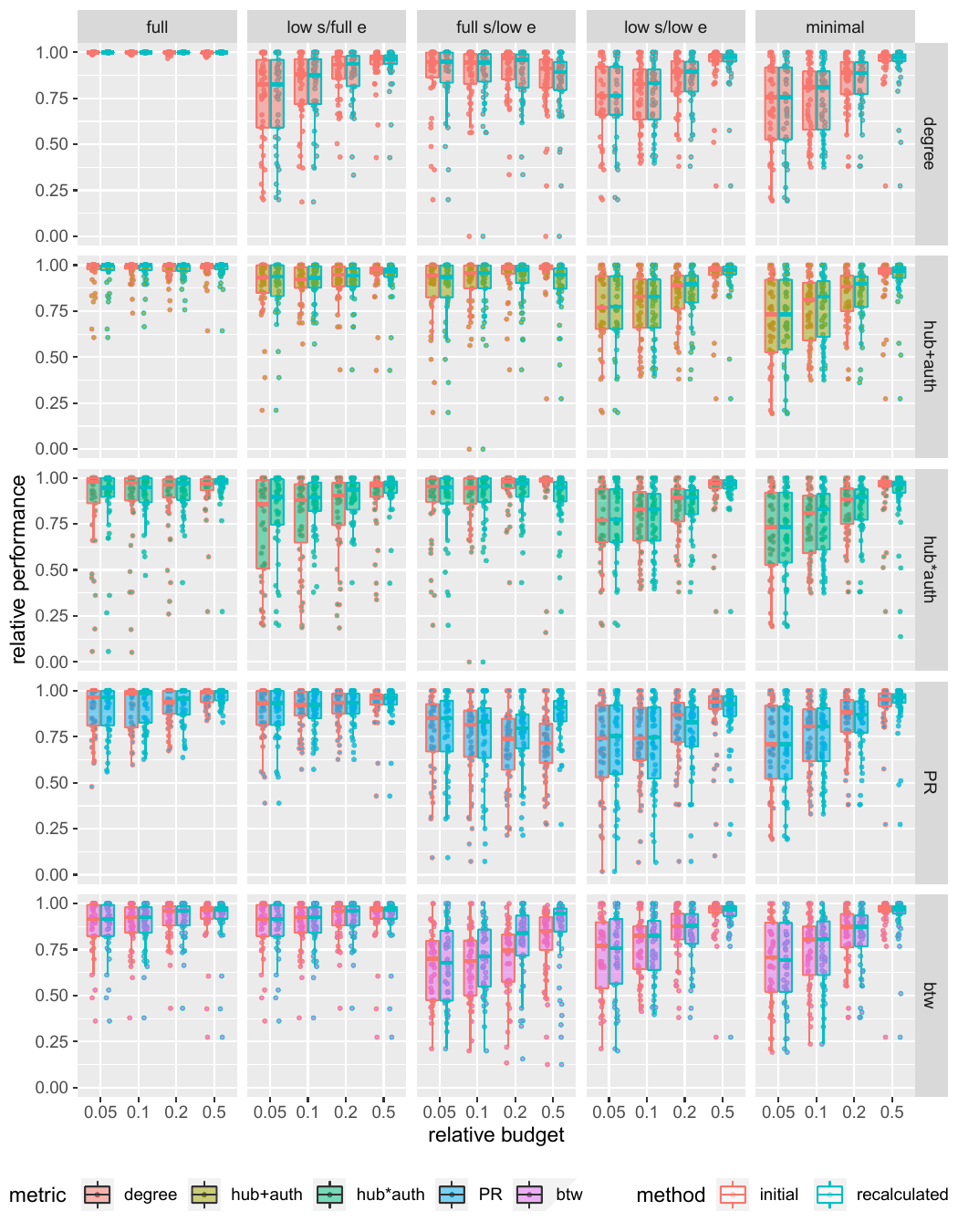}

\caption{Relative performance of initial state vs recalculated strategies for all information levels and all metrics for select relative budgets.}\label{si:fig:recalc}
\end{figure*}

%%%%%%%%%%%%%%%%%%%%

\subsection*{Tables}

\begin{longtable}[]{cccccccc}
\caption{\textbf{Summary statistics of county-level networks.} Infecting trips are those from an infected to uninfected lake; proportion infecting edges are edges with an origin at an infecting lake and destination at an uninfected lake. The 84 counties with at least one lake are listed; our analysis focuses on the counties with at least 10 lakes (the first 58 counties listed).}\label{si:tab:data}
\tabularnewline
\toprule
\multicolumn{1}{|p{1.5cm}|}{\centering County}
& \multicolumn{1}{|p{1cm}|}{\centering Lakes}
& \multicolumn{1}{|p{1.5cm}|}{\centering Infected}
& \multicolumn{1}{|p{2cm}|}{\centering Proportion \\ infected}
& \multicolumn{1}{|p{1cm}|}{\centering Edges}
& \multicolumn{1}{|p{2cm}|}{\centering Trips}
& \multicolumn{1}{|p{1.75cm}|}{\centering Proportion \\
infecting trips}
& \multicolumn{1}{|p{1.75cm}|}{\centering Proportion \\
infecting edges}
\\
\midrule
\endfirsthead
\toprule
\multicolumn{1}{|p{1.5cm}|}{\centering County}
& \multicolumn{1}{|p{1cm}|}{\centering Lakes}
& \multicolumn{1}{|p{1.5cm}|}{\centering Infected}
& \multicolumn{1}{|p{2cm}|}{\centering Proportion \\ infected}
& \multicolumn{1}{|p{1cm}|}{\centering Edges}
& \multicolumn{1}{|p{2cm}|}{\centering Trips}
& \multicolumn{1}{|p{1.75cm}|}{\centering Proportion \\
infecting trips}
& \multicolumn{1}{|p{1.75cm}|}{\centering Proportion \\
infecting edges}
\\
\midrule
\endhead
Itasca & 854 & 7 & 0.01 & 17954 & 183194.95 & 0.08 & 0.06 \\
Otter Tail & 834 & 39 & 0.05 & 12350 & 206542.70 & 0.08 & 0.22 \\
Cook & 759 & 2 & 0.00 & 5531 & 182179.20 & 0.02 & 0.06 \\
Lake & 726 & 2 & 0.00 & 6815 & 171210.15 & 0.02 & 0.04 \\
Saint Louis & 725 & 2 & 0.00 & 19009 & 266655.20 & 0.07 & 0.03 \\
Cass & 529 & 8 & 0.02 & 14613 & 187737.35 & 0.07 & 0.09 \\
Crow Wing & 423 & 30 & 0.07 & 18030 & 151953.00 & 0.09 & 0.20 \\
Becker & 346 & 11 & 0.03 & 9737 & 109867.80 & 0.09 & 0.12 \\
Hubbard & 293 & 2 & 0.01 & 8683 & 120288.30 & 0.08 & 0.03 \\
Beltrami & 253 & 7 & 0.03 & 4773 & 93535.95 & 0.08 & 0.11 \\
Aitkin & 239 & 3 & 0.01 & 3794 & 76799.55 & 0.10 & 0.06 \\
Douglas & 238 & 26 & 0.11 & 5934 & 90715.25 & 0.07 & 0.24 \\
Clearwater & 231 & 2 & 0.01 & 1171 & 58420.35 & 0.04 & 0.07 \\
Stearns & 195 & 6 & 0.03 & 3161 & 66440.25 & 0.11 & 0.10 \\
Wright & 182 & 8 & 0.04 & 6813 & 101686.40 & 0.11 & 0.11 \\
Polk & 168 & 1 & 0.01 & 527 & 26890.15 & 0.06 & 0.08 \\
Pope & 140 & 6 & 0.04 & 1330 & 40824.50 & 0.09 & 0.19 \\
Washington & 133 & 4 & 0.03 & 1521 & 55365.55 & 0.08 & 0.15 \\
Hennepin & 126 & 9 & 0.07 & 3254 & 167916.15 & 0.05 & 0.18 \\
Pine & 123 & 1 & 0.01 & 474 & 23822.15 & 0.09 & 0.08 \\
Kandiyohi & 117 & 10 & 0.09 & 2453 & 54807.30 & 0.11 & 0.21 \\
Carlton & 104 & 1 & 0.01 & 503 & 20779.75 & 0.11 & 0.05 \\
Grant & 103 & 4 & 0.04 & 729 & 23444.85 & 0.09 & 0.19 \\
Clay & 99 & 1 & 0.01 & 263 & 14108.20 & 0.03 & 0.08 \\
Morrison & 94 & 4 & 0.04 & 369 & 21464.85 & 0.07 & 0.21 \\
Todd & 92 & 8 & 0.09 & 1561 & 33351.95 & 0.10 & 0.19 \\
Marshall & 81 & 2 & 0.02 & 634 & 16239.25 & 0.07 & 0.11 \\
Chisago & 74 & 2 & 0.03 & 577 & 28639.25 & 0.11 & 0.10 \\
Big Stone & 72 & 1 & 0.01 & 259 & 14447.70 & 0.07 & 0.08 \\
Anoka & 64 & 1 & 0.02 & 1053 & 39858.60 & 0.14 & 0.05 \\
Dakota & 63 & 5 & 0.08 & 283 & 34870.40 & 0.08 & 0.23 \\
Ramsey & 62 & 8 & 0.13 & 786 & 27668.65 & 0.11 & 0.22 \\
Carver & 56 & 7 & 0.12 & 1147 & 32224.75 & 0.12 & 0.18 \\
Sherburne & 54 & 1 & 0.02 & 426 & 16775.65 & 0.14 & 0.07 \\
Meeker & 53 & 2 & 0.04 & 1032 & 25328.40 & 0.15 & 0.07 \\
Stevens & 44 & 2 & 0.05 & 242 & 9147.60 & 0.10 & 0.15 \\
Scott & 36 & 3 & 0.08 & 370 & 18346.30 & 0.12 & 0.19 \\
Mcleod & 35 & 1 & 0.03 & 499 & 11166.05 & 0.20 & 0.04 \\
Kanabec & 32 & 1 & 0.03 & 148 & 9849.40 & 0.12 & 0.10 \\
Isanti & 31 & 2 & 0.06 & 256 & 14615.65 & 0.13 & 0.12 \\
Le Sueur & 30 & 1 & 0.03 & 512 & 18381.00 & 0.16 & 0.05 \\
Rice & 25 & 1 & 0.04 & 257 & 12246.90 & 0.19 & 0.07 \\
Wadena & 23 & 1 & 0.04 & 67 & 6042.75 & 0.13 & 0.15 \\
Koochiching & 22 & 1 & 0.05 & 58 & 5687.75 & 0.09 & 0.16 \\
Mahnomen & 21 & 1 & 0.05 & 185 & 9260.60 & 0.21 & 0.09 \\
Blue Earth & 20 & 1 & 0.05 & 173 & 10902.85 & 0.18 & 0.08 \\
Brown & 19 & 1 & 0.05 & 86 & 4078.65 & 0.17 & 0.12 \\
Cottonwood & 19 & 1 & 0.05 & 72 & 4044.00 & 0.15 & 0.15 \\
Goodhue & 18 & 3 & 0.17 & 36 & 12430.05 & 0.00 & 0.14 \\
Mille Lacs & 18 & 2 & 0.11 & 35 & 14304.90 & 0.03 & 0.26 \\
Murray & 18 & 1 & 0.06 & 80 & 5165.85 & 0.17 & 0.14 \\
Lyon & 17 & 1 & 0.06 & 83 & 3695.85 & 0.23 & 0.14 \\
Swift & 17 & 1 & 0.06 & 51 & 2012.40 & 0.17 & 0.18 \\
Benton & 13 & 1 & 0.08 & 22 & 3232.40 & 0.17 & 0.14 \\
Jackson & 12 & 1 & 0.08 & 76 & 3519.15 & 0.26 & 0.13 \\
Sibley & 12 & 1 & 0.08 & 25 & 3256.20 & 0.16 & 0.16 \\
Lincoln & 11 & 1 & 0.09 & 53 & 5110.05 & 0.29 & 0.15 \\
Waseca & 10 & 1 & 0.10 & 47 & 5542.50 & 0.18 & 0.13 \\
 \cmidrule(l{2em}r{2em}){1-8}
Olmsted & 9 & 2 & 0.22 & 15 & 3542.40 & 0.03 & 0.13 \\
Winona & 9 & 4 & 0.44 & 39 & 8793.15 & 0.04 & 0.21 \\
Yellow Medicine & 9 & 1 & 0.11 & 18 & 1909.20 & 0.13 & 0.17 \\
Lake Of The Woods & 8 & 1 & 0.12 & 8 & 6829.55 & 0.06 & 0.12 \\
Nobles & 8 & 1 & 0.12 & 44 & 4092.40 & 0.19 & 0.14 \\
Renville & 7 & 1 & 0.14 & 23 & 1231.65 & 0.20 & 0.17 \\
Wabasha & 7 & 2 & 0.29 & 10 & 5349.60 & 0.00 & 0.00 \\
Freeborn & 6 & 1 & 0.17 & 8 & 2367.55 & 0.17 & 0.12 \\
Lac Qui Parle & 6 & 2 & 0.33 & 14 & 3627.55 & 0.01 & 0.21 \\
Mower & 6 & 1 & 0.17 & 6 & 988.85 & 0.01 & 0.17 \\
Steele & 6 & 1 & 0.17 & 6 & 933.75 & 0.12 & 0.17 \\
Wilkin & 6 & 2 & 0.33 & 6 & 1060.55 & 0.00 & 0.00 \\
Traverse & 5 & 1 & 0.20 & 12 & 2559.45 & 0.21 & 0.25 \\
Watonwan & 5 & 1 & 0.20 & 15 & 1024.10 & 0.24 & 0.20 \\
Chippewa & 4 & 0 & 0.00 & 4 & 905.95 & 0.00 & 0.00 \\
Faribault & 4 & 1 & 0.25 & 4 & 495.00 & 0.10 & 0.25 \\
Kittson & 3 & 1 & 0.33 & 3 & 617.55 & 0.14 & 0.33 \\
Martin & 3 & 0 & 0.00 & 3 & 596.15 & 0.00 & 0.00 \\
Nicollet & 3 & 0 & 0.00 & 3 & 472.70 & 0.00 & 0.00 \\
Redwood & 3 & 1 & 0.33 & 3 & 369.80 & 0.08 & 0.33 \\
Roseau & 3 & 1 & 0.33 & 3 & 205.00 & 0.30 & 0.33 \\
Dodge & 2 & 1 & 0.50 & 2 & 508.85 & 0.02 & 0.50 \\
Rock & 2 & 0 & 0.00 & 2 & 490.30 & 0.00 & 0.00 \\
Norman & 1 & 0 & 0.00 & 1 & 248.45 & 0.00 & 0.00 \\
Pennington & 1 & 0 & 0.00 & 1 & 209.90 & 0.00 & 0.00 \\
Pipestone & 1 & 0 & 0.00 & 1 & 156.10 & 0.00 & 0.00 \\
\bottomrule
\end{longtable}

\newpage
\begin{longtable}[]{lllrrrrrr}
\caption{\label{si:tab:results}\textbf{Relative performance across information scenarios, specific budgets and metrics.} Budget is measured as a proportion of the total budget needed to inspect all infected boats. See sections \ref{si:info} and \ref{si:metrics} for detail on the information scenarios and each of the metrics respectively. The remaining columns present summary statistics for the distribution of outcomes across counties ($n=58$). Each of these statistics focuses on the performance of the network-based approach, where performance is measured as the number of infected boats inspected relative to the optimum. We report the median, lower quartile, the average for all counties in the lower quartile, the proportion of counties where the precise optimum is achieved (proportion perfect), and the proportion of counties where less than 66\% of the optimum is achieved.}
\tabularnewline
\toprule
\multicolumn{1}{|p{1.25cm}|}{\centering Budget}
& \multicolumn{1}{|p{2cm}|}{\centering Information}
& \multicolumn{1}{|p{1.75cm}|}{\centering Metric}
& \multicolumn{1}{|p{1.5cm}|}{\centering Median}
& \multicolumn{1}{|p{1.5cm}|}{\centering Lower \\ quartile}
& \multicolumn{1}{|p{1.5cm}|}{\centering Average \\ $<25$\%}
& \multicolumn{1}{|p{1.75cm}|}{\centering Proportion \\ perfect}
& \multicolumn{1}{|p{1.75cm}|}{\centering Proportion \\ $<0.66$} 
\\
\midrule
\endfirsthead
\toprule
\multicolumn{1}{|p{1.25cm}|}{\centering Budget}
& \multicolumn{1}{|p{2cm}|}{\centering Information}
& \multicolumn{1}{|p{1.75cm}|}{\centering Metric}
& \multicolumn{1}{|p{1.5cm}|}{\centering Median}
& \multicolumn{1}{|p{1.5cm}|}{\centering Lower \\ quartile}
& \multicolumn{1}{|p{1.5cm}|}{\centering Average \\ $<25$\%}
& \multicolumn{1}{|p{1.75cm}|}{\centering Proportion \\ perfect}
& \multicolumn{1}{|p{1.75cm}|}{\centering Proportion \\ $<0.66$} 
\\
\midrule
\endhead
\endhead
0.1 & full & degree & 1.00 & 1.00 & 0.99 & 0.88 & 0.00\\
0.1 & full & hub+auth & 1.00 & 0.98 & 0.91 & 0.71 & 0.00\\
0.1 & full & hub*auth & 0.98 & 0.88 & 0.60 & 0.41 & 0.12\\
0.1 & full & PR & 0.99 & 0.80 & 0.74 & 0.49 & 0.06\\
0.1 & full & btw & 0.93 & 0.84 & 0.72 & 0.24 & 0.06\\
\addlinespace
0.1 & low s/full e & degree & 0.88 & 0.72 & 0.55 & 0.20 & 0.16\\
0.1 & low s/full e & hub+auth & 0.92 & 0.88 & 0.77 & 0.24 & 0.02\\
0.1 & low s/full e & hub*auth & 0.88 & 0.65 & 0.40 & 0.22 & 0.25\\
0.1 & low s/full e & PR & 0.92 & 0.87 & 0.74 & 0.25 & 0.04\\
0.1 & low s/full e & btw & 0.93 & 0.84 & 0.72 & 0.24 & 0.06\\
\addlinespace
0.1 & full s/low e & degree & 0.94 & 0.83 & 0.65 & 0.24 & 0.10\\
0.1 & full s/low e & hub+auth & 0.96 & 0.87 & 0.67 & 0.31 & 0.08\\
0.1 & full s/low e & hub*auth & 0.95 & 0.86 & 0.67 & 0.29 & 0.10\\
0.1 & full s/low e & PR & 0.82 & 0.64 & 0.40 & 0.20 & 0.31\\
0.1 & full s/low e & btw & 0.69 & 0.50 & 0.42 & 0.04 & 0.45\\
\addlinespace
0.1 & low s/low e & degree & 0.83 & 0.64 & 0.51 & 0.12 & 0.31\\
0.1 & low s/low e & hub+auth & 0.83 & 0.66 & 0.52 & 0.12 & 0.27\\
0.1 & low s/low e & hub*auth & 0.83 & 0.66 & 0.52 & 0.12 & 0.27\\
0.1 & low s/low e & PR & 0.74 & 0.62 & 0.44 & 0.08 & 0.35\\
0.1 & low s/low e & btw & 0.82 & 0.64 & 0.53 & 0.08 & 0.29\\
\addlinespace
0.1 & minimal & degree & 0.81 & 0.58 & 0.48 & 0.10 & 0.35\\
0.1 & minimal & hub+auth & 0.81 & 0.59 & 0.48 & 0.10 & 0.29\\
0.1 & minimal & hub*auth & 0.81 & 0.59 & 0.48 & 0.10 & 0.29\\
0.1 & minimal & PR & 0.81 & 0.62 & 0.48 & 0.06 & 0.31\\
0.1 & minimal & btw & 0.80 & 0.61 & 0.46 & 0.06 & 0.31\\
\addlinespace
0.5 & full & degree & 1.00 & 0.99 & 0.99 & 0.74 & 0.00\\
0.5 & full & hub+auth & 1.00 & 0.98 & 0.93 & 0.66 & 0.02\\
0.5 & full & hub*auth & 0.97 & 0.93 & 0.82 & 0.26 & 0.03\\
0.5 & full & PR & 0.99 & 0.94 & 0.90 & 0.47 & 0.00\\
0.5 & full & btw & 0.97 & 0.92 & 0.80 & 0.10 & 0.05\\
\addlinespace
0.5 & low s/full e & degree & 0.97 & 0.93 & 0.84 & 0.12 & 0.03\\
0.5 & low s/full e & hub+auth & 0.97 & 0.95 & 0.86 & 0.17 & 0.03\\
0.5 & low s/full e & hub*auth & 0.96 & 0.90 & 0.72 & 0.12 & 0.09\\
0.5 & low s/full e & PR & 0.96 & 0.92 & 0.83 & 0.12 & 0.03\\
0.5 & low s/full e & btw & 0.97 & 0.92 & 0.80 & 0.10 & 0.05\\
\addlinespace
0.5 & full s/low e & degree & 0.91 & 0.81 & 0.62 & 0.16 & 0.09\\
0.5 & full s/low e & hub+auth & 0.99 & 0.98 & 0.90 & 0.38 & 0.02\\
0.5 & full s/low e & hub*auth & 0.99 & 0.98 & 0.85 & 0.38 & 0.03\\
0.5 & full s/low e & PR & 0.72 & 0.61 & 0.48 & 0.05 & 0.38\\
0.5 & full s/low e & btw & 0.85 & 0.75 & 0.53 & 0.00 & 0.24\\
\addlinespace
0.5 & low s/low e & degree & 0.98 & 0.95 & 0.81 & 0.17 & 0.05\\
0.5 & low s/low e & hub+auth & 0.97 & 0.95 & 0.80 & 0.17 & 0.05\\
0.5 & low s/low e & hub*auth & 0.97 & 0.95 & 0.80 & 0.17 & 0.05\\
0.5 & low s/low e & PR & 0.94 & 0.90 & 0.69 & 0.09 & 0.10\\
0.5 & low s/low e & btw & 0.97 & 0.95 & 0.83 & 0.19 & 0.02\\
\addlinespace
0.5 & minimal & degree & 0.98 & 0.95 & 0.86 & 0.19 & 0.02\\
0.5 & minimal & hub+auth & 0.97 & 0.95 & 0.79 & 0.16 & 0.07\\
0.5 & minimal & hub*auth & 0.97 & 0.95 & 0.79 & 0.16 & 0.07\\
0.5 & minimal & PR & 0.95 & 0.92 & 0.80 & 0.16 & 0.03\\
0.5 & minimal & btw & 0.97 & 0.96 & 0.84 & 0.19 & 0.02\\
\bottomrule
\end{longtable}

\newpage
\begin{longtable}[]{lllrrrrrr}
\caption{\label{si:tab:results-bin}Relative performance across information scenarios, budget categories and metrics.}
\tabularnewline
\toprule
\multicolumn{1}{|p{1.5cm}|}{\centering Budget \\category}
& \multicolumn{1}{|p{2cm}|}{\centering Information}
& \multicolumn{1}{|p{1.75cm}|}{\centering Metric}
& \multicolumn{1}{|p{1.5cm}|}{\centering Median}
& \multicolumn{1}{|p{1.5cm}|}{\centering Lower \\ quartile}
& \multicolumn{1}{|p{1.5cm}|}{\centering Average \\ $<25$\%}
& \multicolumn{1}{|p{1.75cm}|}{\centering Proportion \\ perfect}
& \multicolumn{1}{|p{1.75cm}|}{\centering Proportion \\ $<0.66$} 
\\
\midrule
\endfirsthead
\toprule
\multicolumn{1}{|p{1.5cm}|}{\centering Budget \\ category}
& \multicolumn{1}{|p{2cm}|}{\centering Information}
& \multicolumn{1}{|p{1.75cm}|}{\centering Metric}
& \multicolumn{1}{|p{1.5cm}|}{\centering Median}
& \multicolumn{1}{|p{1.5cm}|}{\centering Lower \\ quartile}
& \multicolumn{1}{|p{1.5cm}|}{\centering Average \\ $<25$\%}
& \multicolumn{1}{|p{1.75cm}|}{\centering Proportion \\ perfect}
& \multicolumn{1}{|p{1.75cm}|}{\centering Proportion \\ $<0.66$} 
\\
\midrule
\endhead
(0,0.25] & full & degree & 1.00 & 1.00 & 0.99 & 0.87 & 0.00\\
(0,0.25] & full & hub+auth & 1.00 & 0.97 & 0.89 & 0.66 & 0.01\\
(0,0.25] & full & hub*auth & 0.97 & 0.88 & 0.60 & 0.36 & 0.13\\
(0,0.25] & full & PR & 0.96 & 0.83 & 0.74 & 0.45 & 0.05\\
(0,0.25] & full & btw & 0.93 & 0.84 & 0.72 & 0.23 & 0.06\\
\addlinespace
(0,0.25] & low s/full e & degree & 0.89 & 0.73 & 0.52 & 0.20 & 0.20\\
(0,0.25] & low s/full e & hub+auth & 0.93 & 0.88 & 0.74 & 0.24 & 0.05\\
(0,0.25] & low s/full e & hub*auth & 0.89 & 0.63 & 0.38 & 0.21 & 0.28\\
(0,0.25] & low s/full e & PR & 0.93 & 0.86 & 0.73 & 0.24 & 0.05\\
(0,0.25] & low s/full e & btw & 0.93 & 0.84 & 0.72 & 0.23 & 0.06\\
\addlinespace
(0,0.25] & full s/low e & degree & 0.95 & 0.85 & 0.65 & 0.25 & 0.10\\
(0,0.25] & full s/low e & hub+auth & 0.97 & 0.90 & 0.69 & 0.29 & 0.08\\
(0,0.25] & full s/low e & hub*auth & 0.97 & 0.90 & 0.69 & 0.29 & 0.10\\
(0,0.25] & full s/low e & PR & 0.79 & 0.63 & 0.41 & 0.14 & 0.32\\
(0,0.25] & full s/low e & btw & 0.72 & 0.52 & 0.41 & 0.03 & 0.41\\
\addlinespace
(0,0.25] & low s/low e & degree & 0.84 & 0.66 & 0.52 & 0.11 & 0.27\\
(0,0.25] & low s/low e & hub+auth & 0.85 & 0.67 & 0.53 & 0.11 & 0.25\\
(0,0.25] & low s/low e & hub*auth & 0.85 & 0.67 & 0.53 & 0.11 & 0.25\\
(0,0.25] & low s/low e & PR & 0.79 & 0.63 & 0.42 & 0.09 & 0.32\\
(0,0.25] & low s/low e & btw & 0.83 & 0.66 & 0.51 & 0.09 & 0.27\\
\addlinespace
(0,0.25] & minimal & degree & 0.84 & 0.64 & 0.46 & 0.10 & 0.29\\
(0,0.25] & minimal & hub+auth & 0.83 & 0.65 & 0.47 & 0.08 & 0.27\\
(0,0.25] & minimal & hub*auth & 0.83 & 0.65 & 0.47 & 0.08 & 0.27\\
(0,0.25] & minimal & PR & 0.83 & 0.65 & 0.46 & 0.08 & 0.29\\
(0,0.25] & minimal & btw & 0.82 & 0.63 & 0.45 & 0.07 & 0.32\\
\addlinespace
(0.25,1] & full & degree & 1.00 & 0.99 & 0.99 & 0.74 & 0.00\\
(0.25,1] & full & hub+auth & 1.00 & 0.98 & 0.93 & 0.66 & 0.02\\
(0.25,1] & full & hub*auth & 0.97 & 0.93 & 0.82 & 0.26 & 0.03\\
(0.25,1] & full & PR & 0.99 & 0.94 & 0.90 & 0.47 & 0.00\\
(0.25,1] & full & btw & 0.97 & 0.92 & 0.80 & 0.10 & 0.05\\
\addlinespace
(0.25,1] & low s/full e & degree & 0.97 & 0.93 & 0.84 & 0.12 & 0.03\\
(0.25,1] & low s/full e & hub+auth & 0.97 & 0.95 & 0.86 & 0.17 & 0.03\\
(0.25,1] & low s/full e & hub*auth & 0.96 & 0.90 & 0.72 & 0.12 & 0.09\\
(0.25,1] & low s/full e & PR & 0.96 & 0.92 & 0.83 & 0.12 & 0.03\\
(0.25,1] & low s/full e & btw & 0.97 & 0.92 & 0.80 & 0.10 & 0.05\\
\addlinespace
(0.25,1] & full s/low e & degree & 0.91 & 0.81 & 0.62 & 0.16 & 0.09\\
(0.25,1] & full s/low e & hub+auth & 0.99 & 0.98 & 0.90 & 0.38 & 0.02\\
(0.25,1] & full s/low e & hub*auth & 0.99 & 0.98 & 0.85 & 0.38 & 0.03\\
(0.25,1] & full s/low e & PR & 0.72 & 0.61 & 0.48 & 0.05 & 0.38\\
(0.25,1] & full s/low e & btw & 0.85 & 0.75 & 0.53 & 0.00 & 0.24\\
\addlinespace
(0.25,1] & low s/low e & degree & 0.98 & 0.95 & 0.81 & 0.17 & 0.05\\
(0.25,1] & low s/low e & hub+auth & 0.97 & 0.95 & 0.80 & 0.17 & 0.05\\
(0.25,1] & low s/low e & hub*auth & 0.97 & 0.95 & 0.80 & 0.17 & 0.05\\
(0.25,1] & low s/low e & PR & 0.94 & 0.90 & 0.69 & 0.09 & 0.10\\
(0.25,1] & low s/low e & btw & 0.97 & 0.95 & 0.83 & 0.19 & 0.02\\
\addlinespace
(0.25,1] & minimal & degree & 0.98 & 0.95 & 0.86 & 0.19 & 0.02\\
(0.25,1] & minimal & hub+auth & 0.97 & 0.95 & 0.79 & 0.16 & 0.07\\
(0.25,1] & minimal & hub*auth & 0.97 & 0.95 & 0.79 & 0.16 & 0.07\\
(0.25,1] & minimal & PR & 0.95 & 0.92 & 0.80 & 0.16 & 0.03\\
(0.25,1] & minimal & btw & 0.97 & 0.96 & 0.84 & 0.19 & 0.02\\
\bottomrule
\end{longtable}
\clearpage 

\newpage
\begin{table}
\caption{\textbf{Relative performance of random site selection across specific budgets and metrics.} These results are based on random selection of sites as described in the Methods section. The median, the lower quartile, the proportion perfect, the proportion achieving an objective of less than 66\% of the optimal, and one average were calculated. The average we calculated was the average performance of the replicates in the bottom quartile in terms of performance relative to the optimal.}
\label{si:tab:results-random}
\centering
\begin{tabular}[t]{llrrrrr}
\toprule
\multicolumn{1}{|p{1.5cm}|}{\centering Budget}
& \multicolumn{1}{|p{1.75cm}|}{\centering Metric}
& \multicolumn{1}{|p{1.5cm}|}{\centering Median}
& \multicolumn{1}{|p{1.5cm}|}{\centering Lower \\ quartile}
& \multicolumn{1}{|p{1.5cm}|}{\centering Average \\ $<25$\%}
& \multicolumn{1}{|p{1.75cm}|}{\centering Proportion \\ perfect}
& \multicolumn{1}{|p{1.75cm}|}{\centering Proportion \\ $<0.66$} 
\\
\midrule
0.1 & random & 0.14 & 0.08 & 0.06 & 0 & 1\\
0.5 & random & 0.32 & 0.22 & 0.16 & 0 & 1\\
\bottomrule
\end{tabular}
\end{table}

\newpage
\begin{table}
\caption{\textbf{Relative performance of random site selection across budget categories and metrics.} These results are based on random selection of sites as described in the Methods section for the range of budgets specified in the first column. The median, the lower quartile, the proportion perfect, the proportion achieving an objective of less than 66\% of the optimal, and one average were calculated. The average we calculated was the average performance of the replicates in the bottom quartile in terms of performance relative to the optimal.}
\label{si:tab:results-random-bin}
\centering
\begin{tabular}[t]{llrrrrr}
\toprule
\multicolumn{1}{|p{1.5cm}|}{\centering Budget \\ category}
& \multicolumn{1}{|p{1.75cm}|}{\centering Metric}
& \multicolumn{1}{|p{1.5cm}|}{\centering Median}
& \multicolumn{1}{|p{1.5cm}|}{\centering Lower \\ quartile}
& \multicolumn{1}{|p{1.5cm}|}{\centering Average \\ $<25$\%}
& \multicolumn{1}{|p{1.75cm}|}{\centering Proportion \\ perfect}
& \multicolumn{1}{|p{1.75cm}|}{\centering Proportion \\ $<0.66$} 
\\
\midrule
(0,0.25] & random & 0.15 & 0.10 & 0.06 & 0 & 1\\
(0.25,1] & random & 0.32 & 0.22 & 0.16 & 0 & 1\\
\bottomrule
\end{tabular}
\end{table}

\end{document}